\documentclass[usenatbib,usegraphicx,usepsfrag,onecolumn]{mn2e}
\usepackage[english]{babel}
\usepackage{color}
\def\sige{\sigma_{P_G}}

\def\k{{\bf k}}
\def\kp{k_\parallel}
\def\kpr{{\bf k}_\perp}
\def\kpm{{k}_\perp}

\def\th{{\mathbf \theta}}
\def\u{\vec{U}}
\def\x{\vec{x}}

\def\araa{ARAA}

\def\mnras{MNRAS}

\def\aap{A \& A}
\def\apj{ApJ}

\def\apjl{ApJL}
\def\u{{\bf U}} 

\def\V2{V_2}
\def\V2ij{V_{2ij}}
\def\S{{\mathcal S}}
\def\V{\mathcal{V}}
\def\N{{\mathcal N}}

\def\la{\left\langle}

\def\lsim{~\rlap{$<$}{\lower 1.0ex\hbox{$\sim$}}}

\def\gsim{~\rlap{$>$}{\lower 1.0ex\hbox{$\sim$}}}

\usepackage{graphics}
\usepackage{color}
\usepackage{amsmath}
\usepackage{psfrag}
\begin{document}
\date {} 
\title[21-cm Power spectrum estimator] {The  visibility based Tapered Gridded
  Estimator (TGE) for the redshifted 21-cm power spectrum}

\author[S. Choudhuri et al.]{Samir Choudhuri$^{1}$\thanks{Email:samir11@phy.iitkgp.ernet.in}, Somnath Bharadwaj$^{1}$, Suman Chatterjee$^{1}$, Sk. Saiyad Ali$^{2}$ 
\newauthor Nirupam Roy $^{3}$ and Abhik Ghosh$^{4}$ \\
  $^{1}$ Department of Physics,  \& Centre for Theoretical Studies, IIT Kharagpur,  Pin: 721 302, India\\
$^{2}$ Department of Physics,Jadavpur University, Kolkata 700032, India\\
$^{3}$ Department of Physics, Indian Institute of Science, Bangalore 560012, India\\
$^{4}$ Kapteyn Astronomical Institute, PO Box 800, 9700 AV Groningen, The Netherlands}

\maketitle

\begin{abstract}
We present the improved visibility based Tapered Gridded Estimator (TGE) for
the power spectrum of the diffuse sky signal.  The visibilities are
gridded to reduce the computation, and tapered through a convolution
to suppress the contribution from the outer regions of the
telescope's field of view.  The TGE also internally estimates the
noise bias, and  subtracts this out to give an unbiased
estimate of the power spectrum. An earlier version of the 2D TGE for
the angular power spectrum $C_{\ell}$ is improved and then extended to
obtain the 3D TGE for the power spectrum $P(\k)$ of the 21-cm
brightness temperature fluctuations. Analytic formulas are also
presented for predicting the variance of the binned power spectrum.
The estimator and its variance predictions are validated using
simulations of $150 \, {\rm MHz}$ GMRT observations.  We find that the
estimator accurately recovers the input model for the 1D Spherical
Power Spectrum $P(k)$ and the 2D Cylindrical Power Spectrum
$P(k_\perp,k_\parallel)$, and the predicted variance is also in
reasonably good agreement with the simulations.    
\end{abstract} 

\section{Introduction}
Observations of the redshifted neutral hydrogen (HI) 21-cm radiation 
hold the potential of probing a wide range of cosmological and
 astrophysical phenomena over a large  redshift range $0 < z \lsim 200$
\citep{BA5,furla06,morales10,prichard12,mellema13}. There now are several 
ongoing experiments  such as the Donald
C. Backer Precision Array to Probe the Epoch of Reionization
(PAPER{\footnote{http://astro.berkeley.edu/dbacker/eor}},
\citealt{parsons10}), the Low Frequency Array
(LOFAR{\footnote{http://www.lofar.org/}},
\citealt{haarlem,yata13}) and  the Murchison Wide-field Array
(MWA{\footnote{http://www.mwatelescope.org}} \citealt{bowman13,tingay13})
which aim to measure the power spectrum   of the 21-cm radiation from  
the Epoch of Reionization (EoR, $6 \lsim z \lsim 13$). Future telescopes 
like the Square Kilometer Array (SKA1
LOW{\footnote{http://www.skatelescope.org/}}, \citealt{koopmans15}) 
and the Hydrogen Epoch of Reionization Array
(HERA{\footnote{http://reionization.org/}}, \citealt{neben16}) are planned 
to achieve even higher sensitivity for measuring  the EoR 21-cm power spectrum.
Several other upcoming experiments  like the Ooty Wide Field Array (OWFA;
\citealt{prasad,ali14}), the Canadian Hydrogen Intensity Mapping
Experiment (CHIME{\footnote{http://chime.phas.ubc.ca/};
  \citealt{bandura}), the Baryon Acoustic Oscillation Broadband,
  Broad Beam Array (BAOBAB{\footnote{http://bao.berkeley.edu/};
    \citealt{pober13a}) and  the Square Kilometre Array (SKA1 MID;
    \citealt{bull15}) target the post-Reionization 21-cm signal 
($0 <    z \lsim 6$).

Despite the  sensitive new instruments, the main challenge 
still arises from the fact that 
the cosmological 21-cm   signal is buried in astrophysical foregrounds 
which are $4-5$  orders  of magnitude brighter 
 \citep{shaver99,dmat1,santos05, ali,paciga11,ghosh1,ghosh2}.
 A large variety of techniques have been proposed  to overcome 
this problem and estimate the  21-cm power spectrum. 
The different approaches may be broadly   divided into two classes
(1.) Foreground Removal, and (2.) Foreground Avoidance.
 
The idea in Foreground Removal is to model the foregrounds and 
subtract these out either directly from the  data 
(eg. \citealt{ali}) or from the power spectrum estimator after 
correlating the data (eg.  \citealt{ghosh1,ghosh2}). 
 Foreground Removal is a topic of intense current research  
\citep{jelic08,bowman09,paciga11,chapman12,parsons12,liu12,trott1,pober13,paciga13,
parsons14,trott16}.  

Various studies (eg. \citealt{adatta10}) show that  the foreground 
contribution to the Cylindrical Power Spectrum  $P(k_{\perp},k_{\parallel})$ 
is expected to be restricted within  a wedge in the two dimensional (2D) 
$(k_{\perp},k_{\parallel})$ plane. The idea in Foreground Avoidance is to 
avoid the Fourier modes within the foreground wedge and only use the 
uncontaminated  modes outside the wedge to estimate the 21-cm power 
spectrum \citep{vedantham12,thyag13,pober14,liu14a,liu14b,dillon14,dillon15,zali15}.  
In a recent paper \citet{jacob16} have compared several power spectrum
estimation techniques in the context of MWA.

Point sources dominate the low frequency sky  at the
angular scales $\le  4^{\circ}$ \citep{ali} which are relevant 
for  EoR 21-cm power spectrum  with the telescopes  like the GMRT, LOFAR and
the upcoming SKA.   It is difficult to model and subtract the point sources 
which are located at   the periphery of the telescope's field of view (FoV).
The antenna response  deviates from circular symmetry, and is   highly
frequency and time dependent at the outer parts of the telescope's FoV.
 The calibration also differs from the phase center due to ionospheric 
fluctuations. The residual point sources located far away from the phase
centre cause the signal to  oscillates along the frequency direction
\citep{ghosh1,ghosh2}. This poses a severe problem for foreground removal
techniques which assume a smooth behavior of the signal along the frequency direction. Equivalently, these distant point sources reduce the EoR window by
increasing the area under the foreground wedge in $(k_{\perp},k_{\parallel})$
space \citep{thyag15}. In a recent paper, \citet{pober16} showed that
correctly modelling and subtracting the distant point sources are important for
detecting the redshifted 21-cm signal. Point source subtraction is also
important for measuring the angular power spectrum of the diffuse Galactic
synchrotron radiation \citep{bernardi09,ghosh150,iacobelli13}. Apart 
from being an important foreground component for the EoR 21-cm signal, this is also interesting in its own right.

It is possible to suppress the contribution from the outer parts of the
telescope's FoV by tapering the sky response through a suitably
chosen window function. \citet{ghosh2} have analyzed $610 {\rm MHz}$ GMRT
data to show that it is possible to implement the tapering by convolving the
observed visibilities with the Fourier transform of the window function. It is
found that this reduces the amplitude of the oscillation along the frequency
direction. Our earlier work \citet{samir14} (hereafter Paper I) has introduced
the Tapered Gridded Estimator (TGE) which   
places the findings of \citet{ghosh2} on a sound theoretical
footing. Considering observations at a single frequency, the TGE estimates the 
angular power spectrum $C_{\ell}$ of the 2D sky signal directly from the
measured visibilities while simultaneously tapering the sky response. 
As a test-bed for the TGE, Paper I  considers a situation where 
the point sources have been identified and subtracted out so that 
the residual visibilities are dominated by the Galactic synchrotron 
radiation. This has been used to investigate how well the TGE  is  able
to recover   the angular power spectrum of the input model used to simulate 
the Galactic synchrotron emission at $150 \, {\rm MHz}$.  While  most of the 
analysis was  for the   GMRT, simulations for  LOFAR were also considered. 
These investigations show that the TGE  is able to recover  the input model
$C_{\ell}^M$ to a high level of precision provided the baselines have a
uniform $uv$ coverage. For the GMRT, which has a patchy $uv$ coverage,   
the $C_{\ell}$ is  somewhat overestimated using TGE though the excess 
is largely within the $1\sigma$ errors.  This deviation  is found to be reduced 
in a situation with  a more uniform and denser baseline distribution , like
LOFAR. Paper I  also analyzes the effects of gain errors and the $w$-term. 

In a recent paper \citet{samir16} (hereafter Paper II) we have further
developed  the simulations of Paper~I to include the point sources.  We have
used conventional radio astronomical techniques to model and subtract the
point sources from the central region of the primary beam. As detailed in
Paper~II, it is difficult to do the same for the sources which are far away
from the phase center, and these persist as residuals in the visibility
data. We find that these residual point sources dominate the  $C_{\ell}$
estimated at large baselines. We also show that it is possible to suppress the
contribution from these residual sources located at  the periphery of the
FoV by using TGE with a suitably chosen window function.  

Removing the  noise bias  is an important issue for any power spectrum
estimator. As demonstrated in Paper II, the TGE internally estimates the actual
noise bias from the data and subtracts this out to give an unbiased
estimate of the power spectrum.

In the present work we report the progress on two counts. First, our earlier
implementation of the TGE assumed a uniform and dense baseline $uv$ coverage 
to calculate the normalization coefficient which relates  visibility
correlations to the estimated angular power spectrum $C_{\ell}$. We, however, 
found (Paper I) that this leads to an overestimate of $C_{\ell}$ for
instruments like the GMRT which have a sparse and patchy $uv$ coverage. In
Section 2 of this paper we present an improved TGE which overcomes this
problem by using 
simulations to estimate the normalization coefficient. Second, the entire
analysis of Papers I and II has been restricted to observations at a single
frequency wherein the relevant issue is to quantify the 2D angular
fluctuations of the sky signal.  This, however, is inadequate for the three
dimensional (3D) redshifted HI 21-cm signal  where it is necessary to
also simultaneously quantify the fluctuations along the frequency
direction. In Section 3 of this paper we have generalized the TGE to quantify
the 3D 21-cm signal  and estimate the spatial power spectrum of the 21-cm  
brightness temperature fluctuations $P(\k)$.  We discuss two different
binning schemes which respectively yield the  spherically-averaged (1D) power
spectrum $P(k)$ and the cylindrically-averaged (2D) power spectrum
$P(\kpm,\kp)$, and present theoretical expressions for predicting the expected
variance.  We have validated the estimator and its variance predictions using
simulations which are described in Section 4 and for which the results are
presented in Section 5.  Sections 6 presents the summary and conclusions.

In this paper, we have used cosmological parameters from the (Planck +
WMAP) best-fit $\Lambda$CDM cosmology (\citealt{ade15}).

\section{$C_{\ell}$ estimation}
\label{v2ps}

\subsection{An Improved TGE}
In this section we restrict our attention  to a single  frequency channel $\nu_a$ which we do not  show explicitly in any of the subsequent equations.  The measured visibilities $\V_i$   can be decomposed into two contributions, 
\begin{equation}
\V_i=\S(\u_i)+\N_i
\label{eq:a1}
\end{equation}
the  sky signal and system noise respectively, and $\u_i$ is the baseline corresponding to the  $i$-th   visibility. 
The signal contribution $\S(\u_i)$ records the
Fourier transform of the product of the telescope's primary beam pattern
${\mathcal A}(\th)$ and the specific intensity fluctuation on the sky $\delta I(\th)$. Expressing the signal in terms of brightness
 temperature fluctuations $\delta T(\th)$ we have 
\begin{equation}
\S(\u_i)= \left(\frac{\partial B}{\partial T}\right)  \int d^2 \theta \,  e^{2 \pi i \u_i
 \cdot \th} {\mathcal A}(\th) \delta T(\th),
\label{eq:a2a}
\end{equation}
where 
$B=2k_B T/\lambda^2$  is the Planck function in the Raleigh-Jeans limit which
is valid at the 
frequencies of our interest. In terms of Fourier components we have 
\begin{equation}
\S(\u_i)= \left( \frac{\partial B}{\partial T} \right)  \int \, d^2 U  \,
  \tilde{a}\left(\u_i - \u\right)\, 
 \, \Delta \tilde{T}(\u),   
\label{eq:a2}
\end{equation}
where $\Delta \tilde{ T}(\u)$ and $\tilde{a}\,(\u)$ are the
Fourier transforms of $\delta T(\th) $ and ${\cal A}(\th)$
respectively.  Here we assume that  $\delta T(\th) $ is a
particular realization of a statistically homogeneous and isotropic
Gaussian random process on the sky. Its  statistical properties  are 
completely characterized by the angular power spectrum of the
 brightness temperature fluctuations $C_{\ell}$ defined  through 
\begin{equation}
 \langle \Delta \tilde{T}(\u ) \Delta \tilde{T}^{*}(\u') \rangle = 
\delta_{D}^{2}(\u-\u') \, C_{2 \pi U}\
\label{eq:a3}
\end{equation}
where $\delta_{D}^{2}(\u-\u')$ is a two dimensional Dirac delta
function and $2 \pi U= \ell$, is the angular multipole. The angular brackets $\langle ... \rangle$ here denote an
ensemble average over different realizations of the stochastic
temperature fluctuations on the sky.

The noise  in the different visibilities is uncorrelated,
and we have  
\begin{equation}
\langle \V_i \V_j \rangle = \langle \S_i \S_j \rangle + \langle \mid \N_i \mid^2  \rangle  \delta_{i,j}
\label{eq:a4}
\end{equation}

where  $\langle \mid \N_i \mid^2  \rangle$ is the noise variance of the visibilities, $\delta_{i,j}$ is a 
Kronecker delta and 
\begin{equation}
\langle \S_i \S_j \rangle =
 \left( \frac{\partial B}{\partial T} \right)^2
 \int d^2 U \, \tilde{a}(\u_i-\u) \, \tilde{a}^{*}(\u_j-\u) \, C_{2 \pi U_i} \,
\label{eq:a5}
\end{equation}
This  convolution can be approximated by  a multiplicating factor if $C_{2 \pi U}$ is nearly constant across the
width of   $\tilde{a}(\u_i-\u) $,  which is the situation at large baselines where the antenna separation is large 
compared to the telescope diameter (Paper I), and we have
\begin{equation}
\langle \mid \V_i \mid^2 \rangle =V_0 C \, _{2 \pi U_i} +  \langle \mid \N_i \mid^2  \rangle \,
\label{eq:a6}
\end{equation}
where 
\begin{equation}
V_0= \left( \frac{\partial B}{\partial T} \right)^2
 \int d^2 U \, \mid \tilde{a}(\u_i-\u) \mid^2 \,.
\label{eq:a6a}
\end{equation}
We see that the correlation of a visibility with itself provides an estimate of the angular power 
spectrum, except for the terms $ \langle \mid \N_i \mid^2  \rangle$ which introduce  a positive noise bias.  

It is possible to control the sidelobe response of the telescope's beam patter 
${\mathcal A}(\th)$
by  tapering the sky response
through a frequency independent window function ${\cal W}(\theta)$.  
In this work we use a Gaussian ${\cal W}(\theta)=e^{-\theta^2/\theta^2_w}$  
with $\theta_w$ chosen  so that the window function  cuts off the sky response 
well  before the first null of ${\mathcal A}(\th)$.  This tapering is
achieved by convolving  the measured visibilities with the Fourier 
transform of ${\cal W}(\theta)$.  
We choose a rectangular grid
in the $uv$ plane and consider the convolved visibilities  
\begin{equation}
\V_{cg} = \sum_{i}\tilde{w}(\u_g-\u_i) \, \V_i
\label{eq:a7}
\end{equation}
where  $\tilde{w}(\u)=\pi\theta_w^2e^{-\pi^2U^2\theta_w^2}$ is the Fourier 
transform of ${\cal W}(\theta)$ and $\u_g$ refers to the different grid points.  As shown in Paper I, gridding reduces the computation in comparison to an estimator that uses pairs of visibilities to estimate the power spectrum. We now focus our attention 
on $\S_{cg}$  which is the sky
signal contribution to $\V_{cg}$. This can be written as 
\begin{equation}
\S_{cg}= \left( \frac{\partial B}{\partial T} \right)  \int \, d^2 U  \, 
 \tilde{K}\left(\u_g - \u\right)\,  \, \Delta \tilde{T}(\u),  
\label{eq:a8}
\end{equation}
where 
\begin{equation}
\tilde{K}\left(\u_g - \u\right)= \int d^2 U^{'} 
\tilde{w}(\u_g-\u^{'}) B(\u^{'}) \tilde{a}\left(\u^{'} - \u\right) 
\label{eq:a9}
\end{equation}
is an effective  ``gridding kernel'', and 
\begin{equation}
{\rm B}(\u)=\sum_i \delta^2_D(\u-\u_i)
\label{eq:a10}
\end{equation}
 is the  baseline sampling function of the measured visibilities. 

Proceeding in exactly the same way as we did for eq. (\ref{eq:a6}) we have 
\begin{equation}
\langle \mid \V_{cg} \mid^2 \rangle =V_{1g} C_{2 \pi U_g} + 
\sum_i \mid \tilde{w}(\u_g-\u_i) \mid^2 \langle \mid \N_i \mid^2  \rangle \, , 
\label{eq:a11}
\end{equation}
where 
\begin{equation}
V_{1g}= \left( \frac{\partial B}{\partial T} \right)^2
 \int d^2 U \, \mid \tilde{K}(\u_i-\u) \mid^2 \,.
\label{eq:a10a}
\end{equation}
Here again we see that the correlation of the tapered gridded visibility with
itself provides an estimate of the angular power  
spectrum, except for the terms $ \langle \mid \N_i \mid^2  \rangle$ which
introduces  a positive noise bias.   

Combining equations (\ref{eq:a6}) and (\ref{eq:a11}) we have 

\begin{equation}
\langle \left( \mid \V_{cg} \mid^2 -\sum_i \mid \tilde{w}(\u_g-\u_i) \mid^2
\mid \V_i \mid^2 \right)   \rangle 
=M_g C_{2 \pi U_g}
\label{eq:a11a}
\end{equation}
where 
\begin{equation}
M_g=V_{1g} - \sum_i \mid \tilde{w}(\u_g-\u_i) \mid^2 V_0
\label{eq:a12}
\end{equation}

This allows us to define the Tapered Gridded Estimator (TGE) as 
\begin{equation}
{\hat E}_g= M_g^{-1} \, \left( \mid \V_{cg} \mid^2 -\sum_i \mid
\tilde{w}(\u_g-\u_i) \mid^2  \mid \V_i \mid^2 \right) \,. 
\label{eq:a13}
\end{equation}

The TGE defined here (eq. \ref{eq:a13}) incorporates three novel
  features which are highlighted below. First, the estimator uses the gridded 
visibilities to estimate $C_{\ell}$, this is computationally much faster
 than individually correlating the visibilities. Second, the correlation
 of the gridded visibilities is used to estimate $C_{\ell}$. A 
positive   noise bias is removed by subtracting 
the auto-correlation of the  visibilities. Third, the estimator 
allows us to taper the FoV so as to restrict the contribution from the 
sources in the outer regions and the sidelobes.
 It is, however, necessary to 
note that this comes at a cost which we now discuss. First,  we lose 
information  at the largest angular scales due to the reduced FoV. 
This restricts the smallest $\ell$ value at which it is possible to estimate
the power spectrum. Second,  the reduced FoV results in a larger cosmic 
variance for the smaller angular modes which are within the tapered FoV.

The TGE provides an unbiased estimate of $C_{\ell_g}$ at the angular multipole
$\ell_g=2 \pi U_g$ {\it i.e.} 
\begin{equation}
\langle {\hat E}_g \rangle = C_{\ell_g}
\label{eq:a14}
\end{equation}

 We  use this to define the binned Tapered Gridded Estimator for bin $a$ 
\begin{equation}
{\hat E}_G(a) = \frac{\sum_g w_g  {\hat E}_g}{\sum_g w_g } \,.
\label{eq:a15}
\end{equation}
where $w_g$ refers to the weight assigned to the contribution from any particular 
grid point. In the entire subsequent analysis we have used the weight
$w_g=1$  which assigns equal weightage to all the  grid points which are sampled 
by the baselines.

The binned estimator  has an expectation value 
\begin{equation}
\bar{C}_{\bar{\ell}_a}  = \frac{ \sum_g w_g C_{\ell_g}}{ \sum_g w_g}
\label{eq:ga16}
\end{equation}
where $ \bar{C}_{\bar{\ell}_a}$ is the average  angular power spectrum  at 
 \begin{equation}
\bar{\ell}_a =
\frac{ \sum_g w_g \ell_g}{ \sum_g w_g}
\label{eq:a18}
\end{equation}
which is the   effective angular multipole  for bin $a$. 

\subsection{Calculating $M_g$}
The discussion, till now, has not addressed  how to calculate  $M_g$ which  
is the normalization constant for the TGE (eq. \ref{eq:a13}).  The values of
$M_g$ (eq. \ref{eq:a12}) depend on the baseline distribution
(eq. \ref{eq:a10}) and the form of the tapering  function
${\cal W}(\theta)$, and it is necessary to calculate $M_g$ at every grid
point in the $uv$ plane. 
Our earlier work (Paper I)  presents an analytic  approximation using
which it is possible to estimate $M_g$. 
While this has been found to work very well in a situation where the baselines
have a nearly uniform and dense $uv$ coverage (Fig. 7 of Paper I),  
it leads to an overestimate of $C_{\ell}$ if we have a sparse and non-uniform 
$uv$ coverage. Here we present a different  method to estimate   
$M_g$ which, as we show later, works very well even if we have a   sparse and
non-uniform  $uv$ coverage. 

 We proceed by   calculating simulated  visibilities $[\V_i]_{\rm UAPS}$
 corresponding to an unit angular power spectrum  (UAPS) which has
 $C_{\ell}=1$  
with exactly  the same baseline distribution as the actual observed visibilities. 
We then have (eq. \ref{eq:a11a}) 
\begin{equation}
M_g=\langle \left( \mid \V_{cg} \mid^2 -\sum_i \mid \tilde{w}(\u_g-\u_i) \mid^2 \langle \mid \V_i \mid^2 \right)  \rangle_{\rm UPAS} 
\label{eq:m1}
\end{equation}
which allows us to estimate $M_g$. We average over $N_u$ independent
realizations of the UPAS to reduce the statistical uncertainty
$(\delta M_g/M_g \sim 1/\sqrt{N_u})$ in the estimated $M_g$.

\begin{figure*}
\begin{center}
\psfrag{x}[t][t][1.5][0]{$u$ ($\lambda$)}
\psfrag{y}[b][c][1.5][0]{$v$ ($\lambda$)}
\psfrag{mg}[t][t][1.][0]{log(Mg)}
\includegraphics[width=80mm,angle=-90]{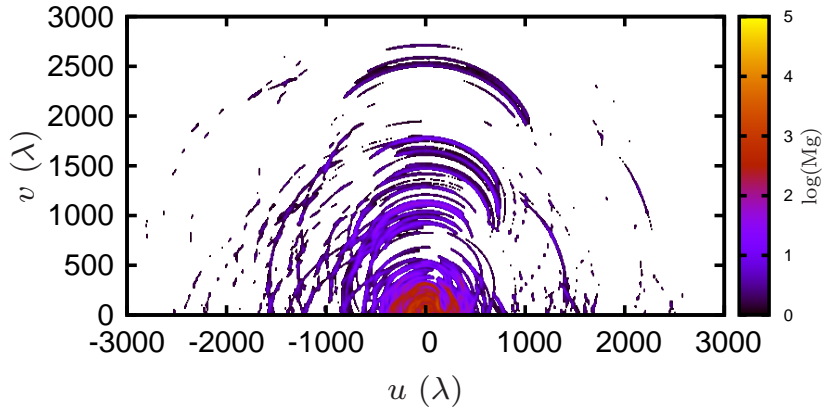}
\caption{ This shows $M_g$ for a fixed value of $f=0.6$. Note that, the baselines in the lower half of the $uv$ plane have been folded on to the upper half.}
\label{fig:fig01}
\end{center}
\end{figure*}

\subsection{Validating the estimator}
\label{sec:clvar}
We have tested the entire method of analysis using simulations of $8$ hours of 
$150 \, {\rm MHz}$ GMRT observations targeted on an arbitrarily selected  field
 located at RA=$10{\rm h} \, 46{\rm  m} \, 00{\rm s}$ and DEC=$59^{\circ} \, 00^{'} \, 59^{''}$.  
The simulations only incorporate the  diffuse Galactic
 synchrotron radiation  for which we  use the measured 
angular power spectrum  \citep{ghosh150} 

\begin{equation}
C^M_{\ell}=A_{\rm 150}\times\left(\frac{1000}{\ell} \right)^{\beta}  \,
\label{eq:cl150}
\end{equation}
as the input model to generate  the brightness temperature fluctuations  on
the sky. Here   $A_{\rm 150}=513 \, {\rm  mK}^2$ and  $\beta=2.34 $ 
\citep{ghosh150}.   The simulation covers a  $\sim 26.4^{\circ} \times 
26.4^{\circ}$ region of the sky,  which is slightly more than ten times  the
FWHM of the GMRT primary beam $(\theta_{FWHM}=157^{'})$. 
The diffuse signal was simulated on a grid of resolution $\sim 0.5^{'}$, and
the entire analysis was restricted to baselines within  $U \le 3,000$. 
Our earlier work (Paper II), and also the discussion of this paper,  show
that the noise bias cancels out from the TGE, and  we have not
included the system noise in these simulations.   

We have modelled the tapering  window function as a 
Gaussian  ${\cal W}(\theta)=e^{-\theta^2/\theta^2_w}$ where we parametrize
$\theta_w=f \theta_0$ where  $\theta_0=0.6 \times \theta_{FWHM}$, 
and preferably $f \le 1$ so that ${\cal W}(\theta)$  cuts off the sky
response well before the first null of the  primary beam. After tapering, we
have an effective  beam pattern ${\mathcal A_W}(\th)={\cal W}(\theta)\,{\cal A}(\th, \nu)$ which is well approximated by a Gaussian $ {\cal A_W}(\theta)=e^{-\theta^2/\theta_1^2}$ with $\theta_1=f (1+f^2)^{-1/2} \theta_0$. The spacing of the  $uv$ grid required for TGE is
decided by $\tilde{a}_W(U)=\pi \theta_1^2 e^{-\pi^2 U^2 \theta_1^2} $ which is
the Fourier transform of ${\mathcal A_W}(\theta)$.  We have chosen a grid
spacing  $\Delta U=\sqrt{\ln2}/(2\pi\theta_1)$ which corresponds to one fourth
of the FWHM of $\tilde{a}_W(U)$.  The convolution in eq. (\ref{eq:a7})   was
restricted to the visibilities within  a disc of radius $12 \times \Delta U$
around each grid point.  The  function $\tilde{w}(\u_g-\u_i)$  falls of
rapidly and we do not expect the visibilities  beyond this to make a
significant contribution.  

We have considered three different values $f=10, 2$ and $0.6$ for the
tapering, here $f=10$ essentially corresponds to a situation with no
tapering, and the sky  response gets confined to a progressively smaller 
region  as the value of $f$ is reduced to $f=2.0$ and $0.6$ respectively
(see Figure 1 of Paper II).  We have
used $N_u=128$ independent realizations of the UAPS to 
estimate $M_g$ for each point in the $uv$ grid. It is necessary to separately
calculate $M_g$ for each value of $f$. 
 Figure \ref {fig:fig01} shows the values of $M_g$ for $f=0.6$. 
We see that this roughly traces out the $uv$ tracks of the baselines, 
the convolution with  $\tilde{w}(\u_g-\u_i)$ results in a thickening 
of the tracks. The values of $M_g$ are roughly proportional to 
$N_g^2 -N_g$, where $N_g$ is the number of visibilities that contribute 
to any particular grid point.

The estimator (eq. \ref{eq:a13}) was applied to the simulated
visibility data which was generated using the model angular power
spectrum (eq. \ref{eq:cl150}).  The estimated angular power spectrum was
binned into $20$ annular bins of equal logarithmic spacing.  We have
used $N_r=128$ independent realizations of the simulation to calculate
the mean and standard deviation of $C_{\ell}$ shown in the left panel
of Figure \ref {fig:fig1}. We see that the TGE is able to recover the
input model $C^M_{\ell}$ quite accurately.  As mentioned earlier, our
previous implementation of TGE (Paper I) had a problem in that the
estimated $C_{\ell}$ was in all cases in excess of the input model
$C^M_{\ell}$, though the deviations were within the $1\sigma$ error
bars throughout. The right panel of Figure \ref {fig:fig1} shows the
fractional deviation $(C_{\ell}-C^M_{\ell}) /C^M_{\ell}$ for the
improved TGE introduced in this paper for the three different values
of $f$ mentioned earlier. We see that for all the values of $f$ the
fractional deviation is less than $10\%$ for $\ell\ge500$.  This is a
considerable improvement over the results of Paper I where we had
$20\%$ to $50\%$ deviations. The fractional deviation is seen to
increase as we increase the tapering {\em i.e.} reduce the value of
$f$. We see that for $f=10$ and $2$, the fractional deviation is less
than $3\%$ for all values of $\ell$ except at the smallest bin.  The fractional deviation for
$f=0.6$ is less than $5\%$ except at the smallest value of  $\ell$  where
it becomes almost $40\%$. This is possibly an outcome of the fact that the 
width of the convolution window $\tilde{w}(\u_g-\u_i)$ increases  as the
value of $f$  is reduced, and the variation of the signal amplitude within
the width of $\tilde{w}(\u_g-\u_i)$ becomes important at small baselines 
where it is reflected as an overestimate of the value of $C_{\ell}$.
Theoretically, we expect the fractional deviation to have random, statistical
fluctuations of the order  $\sigma_{E_G}/\sqrt{N_r}C^M_{\ell}$, where
$\sigma_{E_G}$ is the standard deviation of the estimated angular power
spectrum. We have shown the statistical fluctuation expected  for $f=0.6$ 
as a shaded region in the right panel of Figure \ref {fig:fig1}. We see that
the fractional deviation is roughly consistent with statistical fluctuations
for $\ell \ge 500$. 

\begin{figure*}
\begin{center}
\psfrag{cl}[b][b][1.5][0]{$\ell (\ell+1) C_{\ell}/2 \pi \, [\rm K^2]$}
\psfrag{l}[c][c][1.5][0]{$\ell$}
\psfrag{model}[r][r][1][0]{Model}
\psfrag{tap10}[r][r][1][0]{f=10}
\psfrag{tap2}[r][r][1][0]{f=2}
\psfrag{tap0.6}[r][r][1][0]{f=0.6}
\psfrag{diff}[c][c][1][0]{$(C_{\ell}-C^M_{\ell}) /C^M_{\ell}$}
\includegraphics[width=80mm,angle=0]{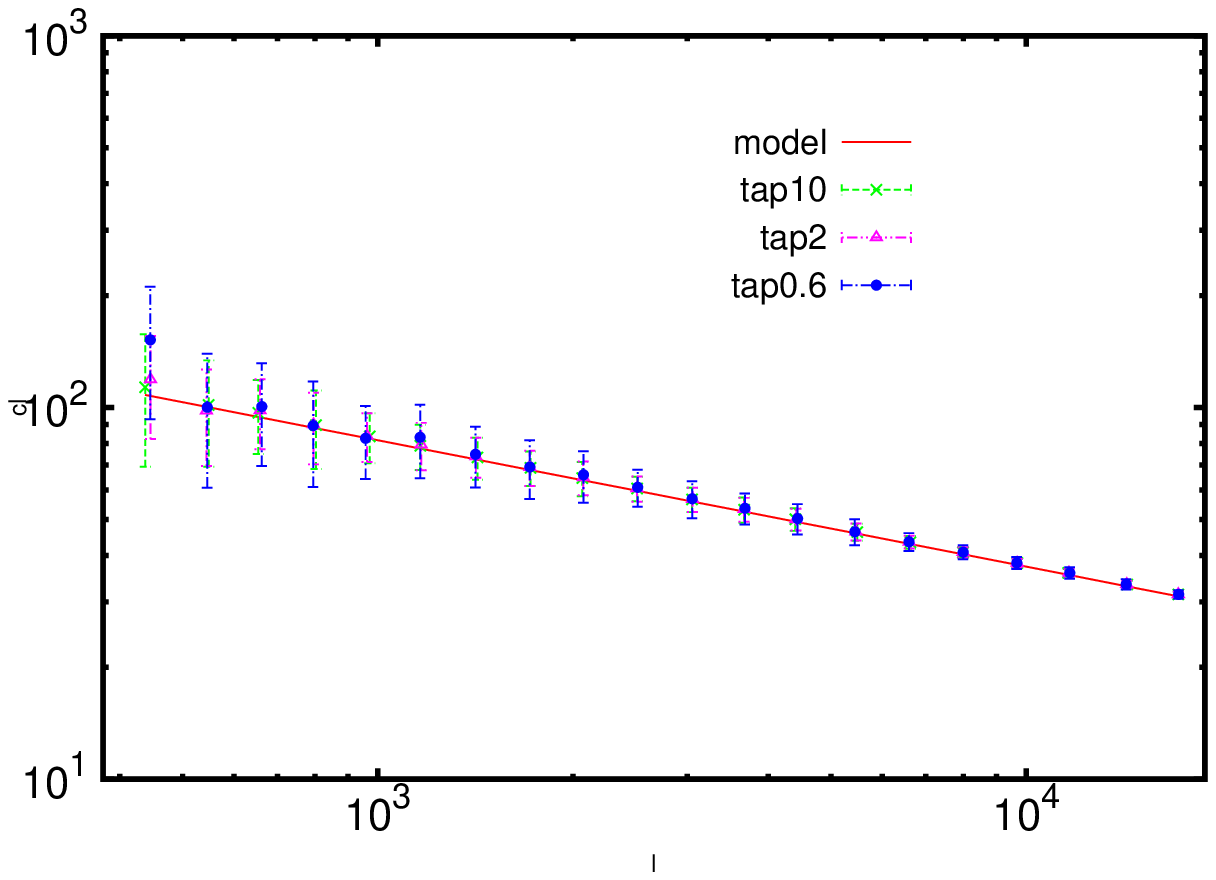}
\includegraphics[width=80mm,angle=0]{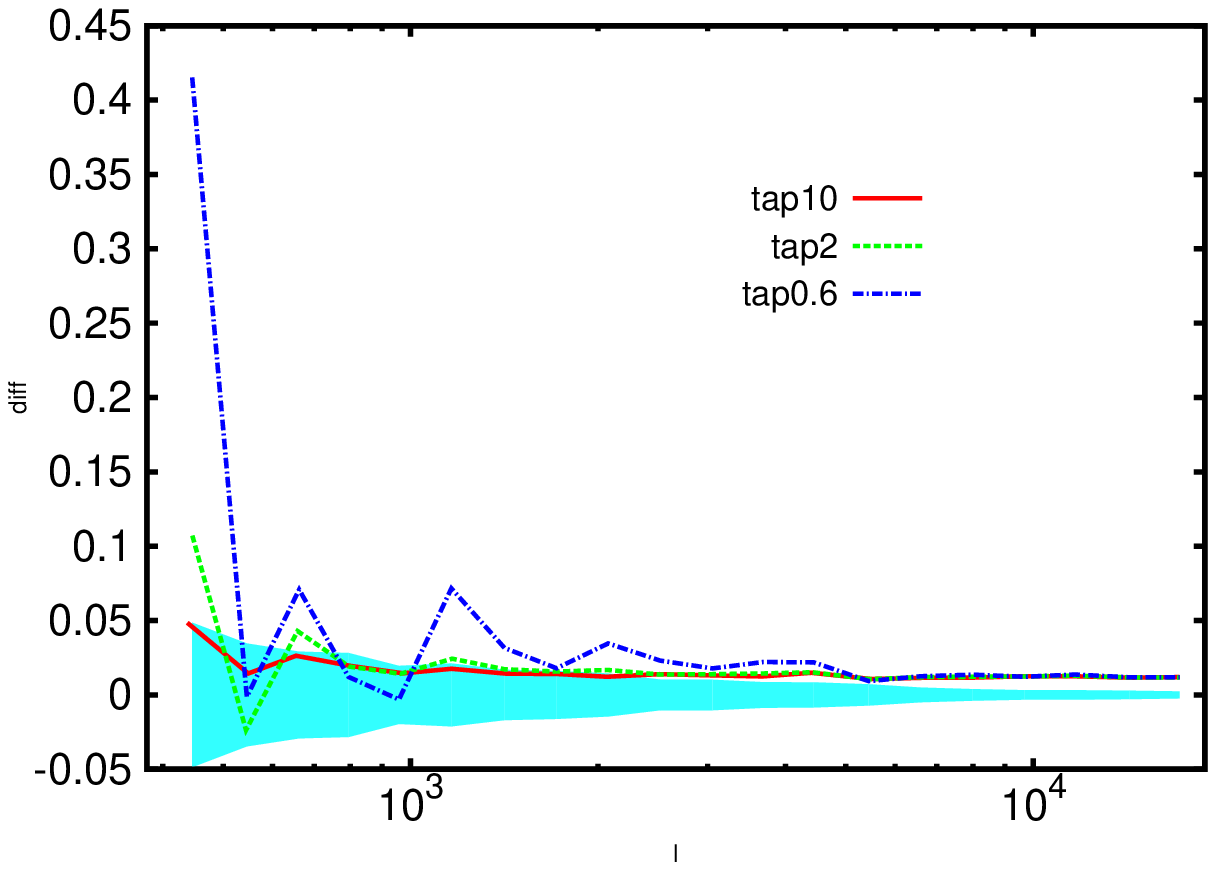}
\caption{The left panel shows a comparison of the input model and the values
  recovered from the simulated visibilities using the improved TGE for
  different tapering of values $f=10,2$ and   $0.6$,  with 1-$\sigma$ error bars
estimated   from   $N_r=128$ realizations of   the simulations. 
 The  right panel shows the fractional deviation of the estimated  $C_{\ell}$
with respect to the   input   model. Here the   shaded region shows 
 the expected statistical fluctuations   ($\sigma_{E_G}/\sqrt{N_r}C^M_{\ell}$)
 of the fractional deviation  for $f=0.6$.} 
\label{fig:fig1}
\end{center}
\end{figure*}

\subsection{Variance}
\label{sec:clvar1}
In the preceding discussion we have used several statistically independent
realizations of the signal to determine the variance of the estimated
binned  angular power spectrum. Such a procedure is, by and large, only
possible with simulated data. We usually   have accessed to only one 
statistically independent realizations  of the input signal, and the aim is to
use this to not only estimate the angular power spectrum but also estimate the 
uncertainty in the estimated angular power spectrum. In this subsection we
present  theoretical predictions for the variance  of the binned TGE
(eq. (\ref{eq:a15})) 
\begin{equation}
\sigma^2_{E_G}(a)=\langle \hat E^2_G(a) \rangle - \langle \hat E_G(a) \rangle^2\,
\label{eq:var1}
\end{equation} 
which can be used to estimate the uncertainty in the measured angular power
spectrum. 
Following Paper I, we ignore the term 
$\sum_i \mid \tilde{w}(\u_g-\u_i) \mid^2\mid \V_i \mid^2$  in
eq. (\ref{eq:a13})  
for calculating the variance. The signal contribution from this term to the 
estimator at the grid point $\u_g$ scales as $N_g$ which is the number of
visibilities that contribute to ${\hat E}_g$. In comparison to this, the
contribution from the term $\mid \V_{cg} \mid^2$ scales as $N_g^2$ which is
much larger when $N_g \gg 1$.  Assuming that this condition is satisfied  at 
every grid point which contributes to the binned TGE, 
it is  justified to drop the term  
$\sum_i \mid \tilde{w}(\u_g-\u_i) \mid^2\mid \V_i \mid^2$  for calculating the
variance. We then have 
\begin{equation}  
\sigma^2_{E_G}(a) = \frac{\sum_{g g^{'}} w_g w_{g^{'}}  M_g^{-1}
  M^{-1}_{g^{'}} \mid \langle \V_{cg}\V^{*}_{cg^{'}} \rangle \mid^2 }{[\sum_{g
    } w_g]^2} \,
\label{eq:var2}
\end{equation}
which is identical to eq. (41) of Paper I, except that we now have  the
normalization constant $M_g^{-1}$ instead  of $K^{-2}_{1g}/V_1$. 

It is necessary to model the correlation between the convolved visibilities at
two different grid points $ \langle \V_{cg}\V^{*}_{cg^{'}} \rangle$
in eq. (\ref{eq:var2})  in order
to make further progress. This correlation is a sum of two parts 
\begin{equation}
\langle \V_{cg}\V^{*}_{cg^{'}} \rangle = 
 \langle \S_{cg}\S^{*}_{cg^{'}} \rangle + 
\langle \N_{cg}\N^{*}_{cg^{'}} \rangle
\label{eq:var2a}
\end{equation}
the  signal  and the  noise correlation respectively. 

 Earlier studies (Paper I) show that we expect the
 signal correlation  $\langle \S_{cg}\S^{*}_{cg^{'}} \rangle$
to fall off as $e^{-\mid \Delta \u_{g g^{'}} \mid^2/\sigma_1^2}$
if  the grid separation is increased, here
$\sigma_1=f^{-1}\sqrt{1+f^2}\sigma_0$ 
where $\sigma_0=0.76/\theta_{\rm FWHM}$. We use this to approximate the 
signal correlation as
\begin{equation}
\langle \S_{cg}\S^{*}_{cg^{'}} \rangle= \sqrt{M_g M_{g^{'}}} e^{-\mid \Delta
  \u_{g g^{'}} \mid^2/\sigma_1^2} \, \bar{C}_{\bar{\ell}_a} 
\label{eq:var3}
\end{equation}
where $\bar{C}_{\bar{\ell}_a} $ refers to the  angular power spectrum
measured at the particular bin $a$ for which the variance $\sigma^2_{E_G}(a)$ 
is being calculated.   

The noise correlation
\begin{equation}
\langle \N_{cg}\N^{*}_{cg^{'}} \rangle= \sum_i  \tilde{w}(\u_g-\u_i)
\tilde{w}^{*}(\u_{g^{'}}-\u_i)  \langle \mid \N_i \mid^2  \rangle \,  
\label{eq:var4}
\end{equation}
also is expected to   fall off as the grid separation is increased, and we
have modeled this $\mid \Delta   \u_{g g^{'}} \mid$ dependence as 
\begin{equation}
\langle \N_{cg}\N^{*}_{cg^{'}} \rangle= \sqrt{K_{2gg}K_{2g^{'}g^{'}}}e^{-
  \mid \Delta \u_{g g^{'}} \mid^2/\sigma_2^2}(2\sigma_n^2)\, 
\label{eq:var5}
\end{equation}
where, $K_{2gg}=\sum_i \mid \tilde{w}(\u_g-\u_i) \mid^2 $, $\sigma_2=3\sigma_0f^{-1}$ and $\sigma_n^2$ is
the variance of the real (and also imaginary) part of $\N_i$. 

We have used eqs. (\ref{eq:var5}), (\ref{eq:var3})  and (\ref{eq:var2a}) in
eq. (\ref{eq:var2}) to calculate $\sigma^2_{E_G}(a)$,  the analytic prediction
for the variance of the estimated  binned angular power spectrum $
\bar{C}_{\bar{\ell}_a}$.

\begin{figure*}
\begin{center}
\psfrag{sig}[b][b][1.2][0]{$\ell (\ell+1) \sigma_{E_G}/2 \pi\, [\rm K^2] $}
\psfrag{l}[c][c][1.5][0]{$\ell$}
\psfrag{sim0.6}[r][r][1][0]{Simulation}
\psfrag{ana0.6}[r][r][1][0]{Analytic}
\psfrag{tap10}[r][r][1][0]{f=10}
\psfrag{tap2}[r][r][1][0]{f=2}
\psfrag{tap0.6}[r][r][1][0]{f=0.6}
\psfrag{f0.6}[r][r][1][0]{f=0.6}
\includegraphics[width=80mm,angle=0]{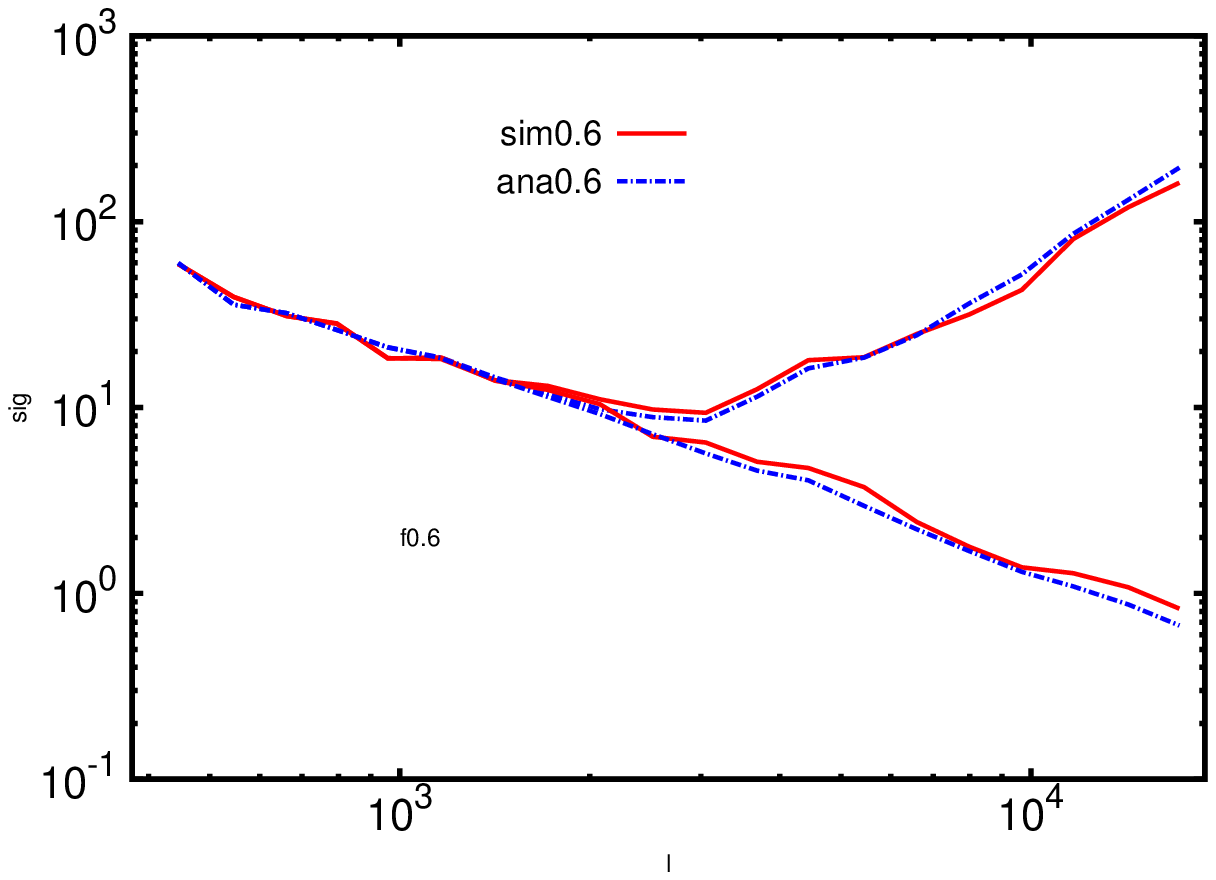}
\includegraphics[width=80mm,angle=0]{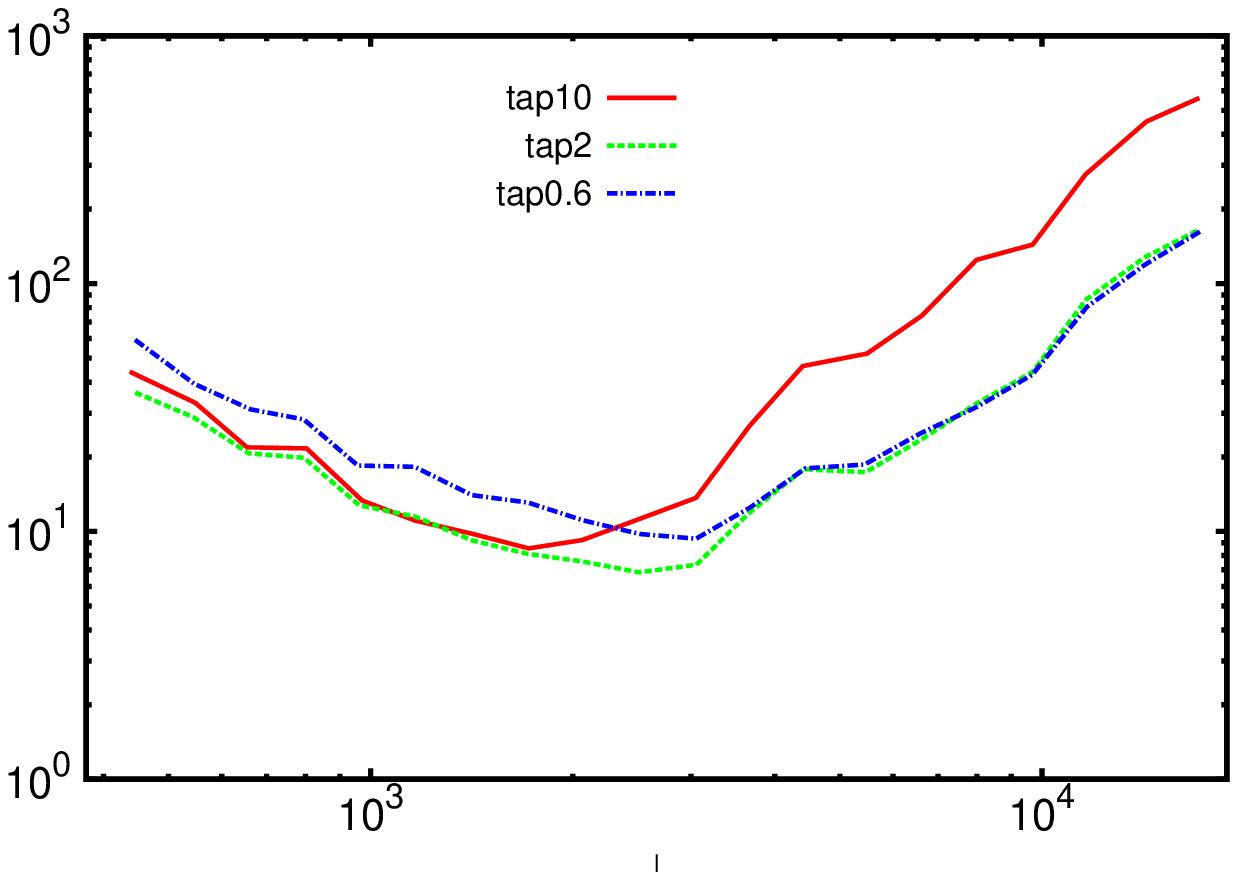}
\caption{In the  left panel  the analytic prediction for the
  variance (eq. \ref{eq:var2}) is compared with variance estimated from
  $N_r=128$ realizations of the simulated visibilities. Results are shown both 
  with (upper curves) and without (lower curves) the system noise contribution. Both match at small $\ell$
  where cosmic variance dominates, the system noise however is important at
  large $\ell$ where the two sets of results are different. 
The right panel shows how the variance with system noise obtained from simulations varies for different values of $f$.}
\label{fig:fig2}
\end{center}
\end{figure*} 

The left panel of Figure \ref {fig:fig2}  shows the analytic prediction for
the variance calculated using eq. (\ref{eq:var2}) for a fixed value of
$f=0.6$. For comparison we also show  the variance 
estimated from $N_r=128$ independent realizations of the simulated
visibilities. We have considered two situations, the first where the simulated
visibilities only have the signal corresponding to the input model
(eq. \ref{eq:cl150}) and no system noise,  and  the second situation where in
addition to the signal the visibilities also have a system noise
contribution  with $\sigma_n=1.03 \, {\rm Jy}$ which corresponds to $16 \, {\rm s}$
integration time and a  channel width of  $125 \, {\rm kHz}$.   
We see that the variance calculated from the simulations is dominated by
cosmic variance at small $\ell$ $(\le 2,000)$ where the
variance  does not  change irrespective of whether we include the system noise
or not. The variance calculated from the simulations is dominated by the  
system noise at large $\ell$ $(\ge 5,000)$. We see that the analytic
predictions  are in reasonably good  agreement with the values
obtained from the simulations over the entire $\ell$ range that we have
considered here. We have also considered  situations where
$f=2.0$ and  $10$ for which the comparison with the analytic results are not
shown here.  In all cases we find that analytic predictions are in
reasonably good agreement with the values obtained from the simulations.  

The right panel of Figure \ref{fig:fig2} shows how the variance obtained from
the simulations changes with $f$. We see that at low $\ell$ the variance
increases  if the value of $f$ is reduced. This is a consequence of the fact
that cosmic variance increases as the sky response is tapered by reducing
$f$. The same effect has also been discussed in detail in our
 earlier paper
(Paper I). We also see that at large $\ell$  the variance is 
considerably higher   
for  $f=10$ in comparison with $f=2$ and $0.6$. This $\ell$ range  is
dominated by the system noise contribution. The number of 
independent visibilities which are combined to estimate the power spectrum at
any grid point increases as $f$ is reduced, and this is reflected in a smaller
variance as $f$ is reduced. 

\section{3D $P(\kpr,\kp)$ estimation}
\subsection{3D TGE}
\label{sec:3dtge}
We now turn our attention to  
the  redshifted $21$-cm HI  brightness temperature fluctuations where it is
necessary to consider  different frequency channels for which 
eq. (\ref{eq:a1}) is generalized to 
\begin{equation}
\V_i(\nu_a)=\S(\u_i,\nu_a)+\N_i(\nu_a).
\label{eq:b1}
\end{equation}
Proceeding in exactly  the same manner as for a single frequency channel
(eq. \ref{eq:a2a}), we have 
\begin{equation}
\S(\u_i,\nu_a)= \left(\frac{\partial B}{\partial T}\right)_{\nu_a}  \int d^2
\theta \,  e^{2 \pi i \u_i  \cdot \th} {\mathcal A}(\th,\nu_a) \delta
T(\th,\nu_a), 
\label{eq:b2a}
\end{equation}
and the noise in the different visibility measurements at different frequency channels
are uncorrelated 
\begin{equation}
\langle \N_i(\nu_a) \N_j(\nu_b) \rangle =  \langle \mid \N_i(\nu_a) \mid^2  \rangle  \delta_{i,j}
\delta_{a,b} \,.
\label{eq:b3}
\end{equation}
Note that the baseline corresponding to a fixed antenna separation
$\u_i={\bf d}_i/\lambda$, the antenna beam pattern  ${\mathcal A}(\th,\nu_a)$ 
 and the factor $\left( \frac{\partial B}{\partial T} \right)_{\nu_a}$ all
vary with the frequency $\nu_a$ in eq. (\ref{eq:b2a}). However, for the present
analysis we only consider the frequency dependence of the HI signal  $\delta T(\th,\nu_a)$ which is assumed to vary much more rapidly with $\nu_a$ in comparison 
to the other terms which are expected to have a relatively slower  frequency
dependence which has been ignored here.  We then have 
\begin{equation}
\S(\u_i,\nu_a)= \left( \frac{\partial B}{\partial T} \right)  \int \, d^2 U  \,
  \tilde{a}\left(\u_i - \u\right)\, 
 \, \Delta \tilde{T}(\u,\nu_a),   
\label{eq:b2}
\end{equation}
which is similar to eq. (\ref{eq:a2}) introduced earlier. 

In eq. (\ref{eq:b2}),  we can express  $\Delta \tilde{T}(\u,\nu)$  in terms of  $\Delta T(\k)$ which refers to the three
dimensional (3D) Fourier decomposition of the HI brightness temperature
fluctuations in the region of space from which the redshifted 21 cm radiation
originated. We use 
equation (7) of  \citet{BS01} (or equivalently eq. (12) of
\citet{BA5}) to express $\S(\u_i,\nu)$ in terms of the three dimensional
brightness temperature fluctuations 
\begin{equation}
\S(\u_i,\nu)= \left( \frac{\partial B}{\partial T} \right)  \int \,
\frac{d^3 k}{(2 \pi)^3}  \,
  \tilde{a}\left(\u_i - \frac{\kpr r}{2 \pi}\right)\, e^{- i \kp r^{'} \nu}
 \, \Delta \tilde{T}(\k),   
\label{eq:b5}
\end{equation}
where $(\kpr,\kp)$ are the components of the comoving wave vector $\k$ respectively
perpendicular and parallel to the line of sight, $r$ is the comoving distance
corresponding to the redshifted 21-cm radiation at the observing frequency
$\nu$, $r^{'}=\mid dr/d \nu \mid$, and 
\begin{equation}
\langle \Delta \tilde{T}(\k) \, \Delta \tilde{T}^{*}(\k^{'}) \rangle = (2
\pi)^3 \delta^3_D(\k-\k^{'}) P(\kpr,\kp)
\label{eq:b6}
\end{equation}
defines $P(\kpr,\kp)$, the 3D power spectrum of HI brightness temperature
fluctuations.   $\nu$ here is measured with respect to the central frequency of
the observation, and   $r$ and $r^{'}$ are held fixed at the values
corresponding to the central frequency.  

We next consider observations with  $N_c$ discrete frequency channels $\nu_a$
with $a=0,1,2,...,N_c-1$, each channel of
width $\Delta \nu_c$ and the total spanning a frequency bandwidth ${\rm
  B_{bw}}$. This corresponds to a comoving spatial extent of  ($r^{'}\rm
  B_{bw}$) along the line of sight and $\kp$ now assumes discrete values 
\begin{equation}
\kp=\frac{2 \pi \tau_m}{r^{'}}
\label{eq:b7}
\end{equation}
where $\tau_m$ is the delay variable (\citealt{morales04,mac06}) which takes values $\tau_m=m/{\rm B_{bw}}$ with
$-N_c/2 < m \le Nc/2$. The $\kp$  integral  in eq.  (\ref{eq:b5})  is now 
replaced   by a discrete sum $\int \kp/(2 \pi) \rightarrow (r^{'}{\rm B_{bw}})^{-1}\sum_m$. It is   further convenient to use 
\begin{equation}
\kpr =\frac{2 \pi \u}{ r}
\label{eq:b8}
\end{equation}
whereby 
\begin{equation}
\S(\u_i,\nu_a)= \left( \frac{\partial B}{\partial T} \right)  \int \,
d^2 U \,  \tilde{a}\left(\u_i - \u \right)\, \sum_m e^{- 2\pi i \tau_m  \nu_a} 
 \, \frac{\Delta \tilde{T}(\u,\tau_m)}{{\rm B_{bw}}\, r^2 r^{'}}\,.   
\label{eq:b9}
\end{equation}
Note here that we can identify  $\tau_m$ as being the Fourier conjugate of
$\nu_a$.  

We now consider the Fourier transform along the frequency axis 
of the measured visibilities which gives the visibilities 
$v_i(\tau_m)$ in delay space
\begin{equation}
v_i(\tau_m)=(\Delta \nu_c) \sum_a e^{2 \pi i \tau_m \nu_a} \,  \V_i(\nu_a) \,.
\label{eq:b11}
\end{equation}
The subsequent analysis of this section is entirely based on the delay space
visibilities $v_i(\tau_m)$ defined in eq. (\ref{eq:b11}). 

Calculating $s(\u_i,\tau_m)$,  the HI signal contribution to $v_i(\tau_m)$ using 
eq. (\ref{eq:b9}),  we have 
\begin{equation}
s(\u_i,\tau_m)= \left( \frac{\partial B}{\partial T} \right)  \int \,
d^2 U\,  \tilde{a}\left(\u_i - \u \right)\, 
 \, \left[ \frac{\Delta \tilde{T}(\u,\tau_m)}{r^2 r^{'}} \right] \,,   
\label{eq:b12}
\end{equation}
where rewriting eq. (\ref{eq:b6}) in terms of the new variables $\u$ and 
$\tau_m$ we have  
\begin{equation}
\langle \Delta \tilde{T}(\u,\tau_m) \,\Delta \tilde{T}^{*}(\u,\tau_n) \rangle
= \delta_D^2(\u-\u^{'}) \left[ \delta_{m,n} ({\rm B_{bw}}\, r^2 r^{'})
  P(\kpr,\kp) \right] \,. 
\label{eq:b13}
\end{equation}
We see that the signal at two different delay channels is uncorrelated. It is
straight forward to also verify that the noise contribution $n_i(\tau_m)$ at
two different delay channels is uncorrelated.  

In summary of the calculations discussed till now in this section, we see that
the visibilities $v_i(\tau_m)$ 
at two  different delay channels are uncorrelated. It therefore suffices to
individually analyze each delay channel separately, and in  the subsequent
discussion  we restrict our attention to a fixed delay channel $\tau_m$.
Calculating  the correlation of a visibility with  itself, we have  
\begin{equation}
\langle \mid v_i(\tau_m) \mid^2  \rangle = V_0 \left[ \frac{{\rm 
      B_{bw}}}{r^2 r^{'}}   P(\kpr,\kp) \right]
 + (\Delta \nu_c)^2  \sum_a \langle \mid \N_i(\nu_a)  \mid^2  \rangle \,. 
\label{eq:b14}
\end{equation}
It is important to note that eqs. (\ref{eq:b12}), (\ref{eq:b13})  and 
(\ref{eq:b14}) which hold for a fixed delay channel
 are exactly analogous  to eqs. (\ref{eq:a2}), (\ref{eq:a3})  and
(\ref{eq:a6})   which hold for  a fixed frequency channel.  We define the 
convolved visibilities in exact analogy with eq. (\ref{eq:a7}) 
\begin{equation}
v_{cg}(\tau_m) = \sum_{i}\tilde{w}(\u_g-\u_i) \, v_i(\tau_m) \,,
\label{eq:b15}
\end{equation}
and we define the 3D TGE  in exact analogy with eq.  (\ref{eq:a13}). 
\begin{equation}
{\hat P}_g(\tau_m)= \left(\frac{M_g {\rm B_{bw}}}{r^2 r^{'}} \right) ^{-1} \,
\left( \mid v_{cg}(\tau_m) \mid^2 -\sum_i \mid \tilde{w}(\u_g-\u_i) \mid^2  
\mid v_i(\tau_m) \mid^2 \right) \,.  
\label{eq:a16}
\end{equation}

The 3D TGE is, by construction,   an unbiased estimator of the three
dimensional power spectrum $P(\kpr,\kp)$, and  we have  
\begin{equation}
\langle {\hat P}_g(\tau_m) \rangle = P({\kpr}_g,{\kp}_m)
\end{equation}
 where ${\kp}_m$ and ${\kpr}_g$ are related to $\tau_m$ and $\u_g$ through eqs.  
(\ref{eq:b7}) and (\ref{eq:b8}) respectively.

\subsection{Frequency Window Function}
\label{filter}
The discrete Fourier transform used to calculate $v_i(\tau_m)$ 
in eq. (\ref{eq:b11}) assumes that the measured visibilities $\V_i(\nu_a)$ are 
periodic across the frequency bandwidth ${\rm B_{bw}}$  ({\em i.e.}
$\V_i(\nu_a)=\V_i(\nu_a+ {\rm B_{bw}})$. In reality, the
measured visibilities are not periodic over the observational bandwidth,
and the discrete Fourier transform encounters 
a discontinuity at the edge of the band. It is possible to avoid
this problem by multiplying the measured visibilities with a frequency window
function 
$F(\nu_a)$ which smoothly falls to zero at the edges of the band.  This
effectively makes the product $F(\nu_a) \times  \V_i(\nu_a)$ periodic,   thereby
doing away with the discontinuity  at the edges of the band. 
This issue has been studied by   \citet{vedantham12} and  \citet{thyag13} 
who have proposed the Blackman-Nuttall \citep{nut81} window function 
\begin{equation}
F(a)=c_0-c_1{\rm cos}\big(\frac{2\pi a}{N_c-1}\big)+c_2{\rm cos}\big(\frac{4\pi a}{N_c-1}\big)-c_3{\rm cos}\big(\frac{6\pi a}{N_c-1}\big) \,
\label{eq:filter}
\end{equation} 
where $c_0=0.3635819,c_1=0.4891775,c_2=0.1365995$ and $c_3=0.0106411$.  In a
recent paper, \citet{chapman14}  
have compared different frequency window functions to  conclude that the
extended Blackman-Nuttall window is 
the best choice for recovering the HI power spectrum. For the present work we
have used the Blackman-Nuttall window as given by eq. (\ref{eq:filter})
above.  The left panel of
Figure \ref{fig:fig3} shows the frequency window function  for $256$ frequency
channels spanning a frequency bandwidth of ${\rm B_{bw}}=16 \, {\rm MHz}$
which corresponds to the values which we have used in our simulations (discussed later).

We now have 
\begin{equation}
v^f_i(\tau_m)=(\Delta \nu_c) \sum_a e^{2 \pi i \tau_m \nu_a} \,  F(\nu_a)
\V_i(\nu_a) \, 
\label{eq:c1}
\end{equation}
where $v^f_i(\tau_m)$ refer to  the delay space visibilities after
introducing the frequency window function.  The filtered delay space
visibilities  
$v^f_i(\tau_m)$ are related to the original delay space visibilities 
$v_i(\tau_m)$ (eq. (\ref{eq:b11})) through a convolution 
\begin{equation}
v^f_i(\tau_m)=\frac{1}{{\rm B_{bw}}} \sum_n   \tilde{f}(\tau_m-\tau_n)v_i(\tau_n) \,
\label{eq:c2}
\end{equation}
where $\tilde{f}(\tau)$ is  the Fourier transform of the frequency window
$F(\nu)$. Recollect that the delay space visibilities  $v_i(\tau_m)$ 
at the different $\tau_m$ are all independent and uncorrelated.
We however  see that this does not hold for the filtered delay space
visibilities $v^f_i(\tau_m)$  for which 
the different $\tau_m$ values are correlated,  the extent of this
correlation being  determined by the width of the function
$\tilde{f}(\tau_m-\tau_n)$ in eq. (\ref {eq:c2}). We now use this to calculate  
the correlation of $v^f_i(\tau_m)$  at two different values of $\tau_m$ for
which we have  
\begin{equation}
\langle v^f_i(\tau_m) v^{f*}_i(\tau_n)   \rangle = \frac{1}{{\rm B_{bw}}^2} \sum_a
\tilde{f}(\tau_m-\tau_a) \tilde{f}^{*}(\tau_n-\tau_a)  \langle 
\mid v_i(\tau_a) \mid^2  \rangle \,. 
\label{eq:c3a}
\end{equation}

This gives the self-correlation to be 
\begin{equation}
\langle \mid v^f_i(\tau_m) \mid^2  \rangle = \frac{1}{{\rm B_{bw}}^2} \sum_a
\mid \tilde{f}(\tau_m-\tau_a) \mid^2 \langle \mid v_i (\tau_a)\mid^2  \rangle \,.
\label{eq:c3}
\end{equation}

\begin{figure*}
\begin{center}
\psfrag{chan}[t][t][1.][0]{Channel Number}
\psfrag{delay}[t][t][1.][0]{Delay channel Number}
\psfrag{htau2}[b][b][1.][0]{$\mid\tilde{f}(\tau)\mid^2$}
\psfrag{hnu}[c][c][1.][0]{$F(\nu)$}
\includegraphics[width=80mm,angle=0]{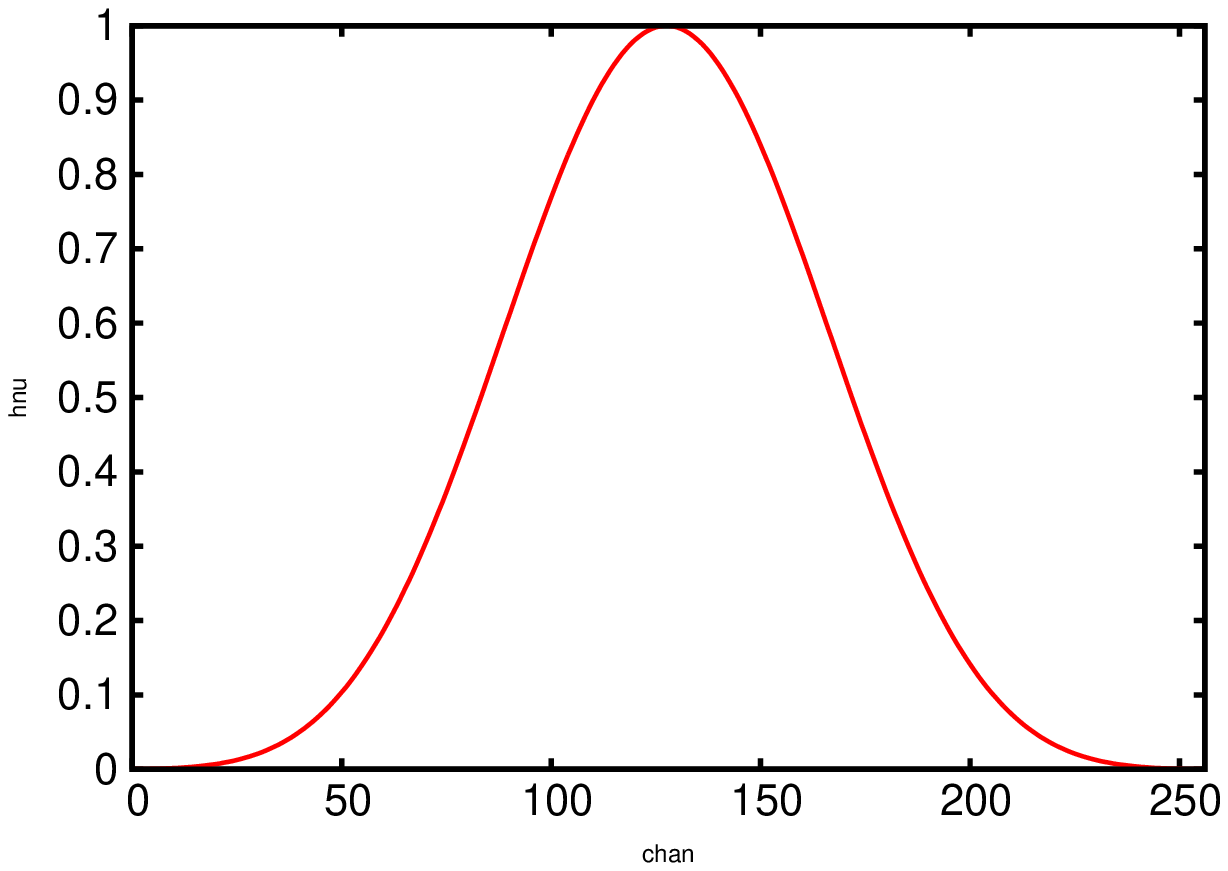}
\includegraphics[width=80mm,angle=0]{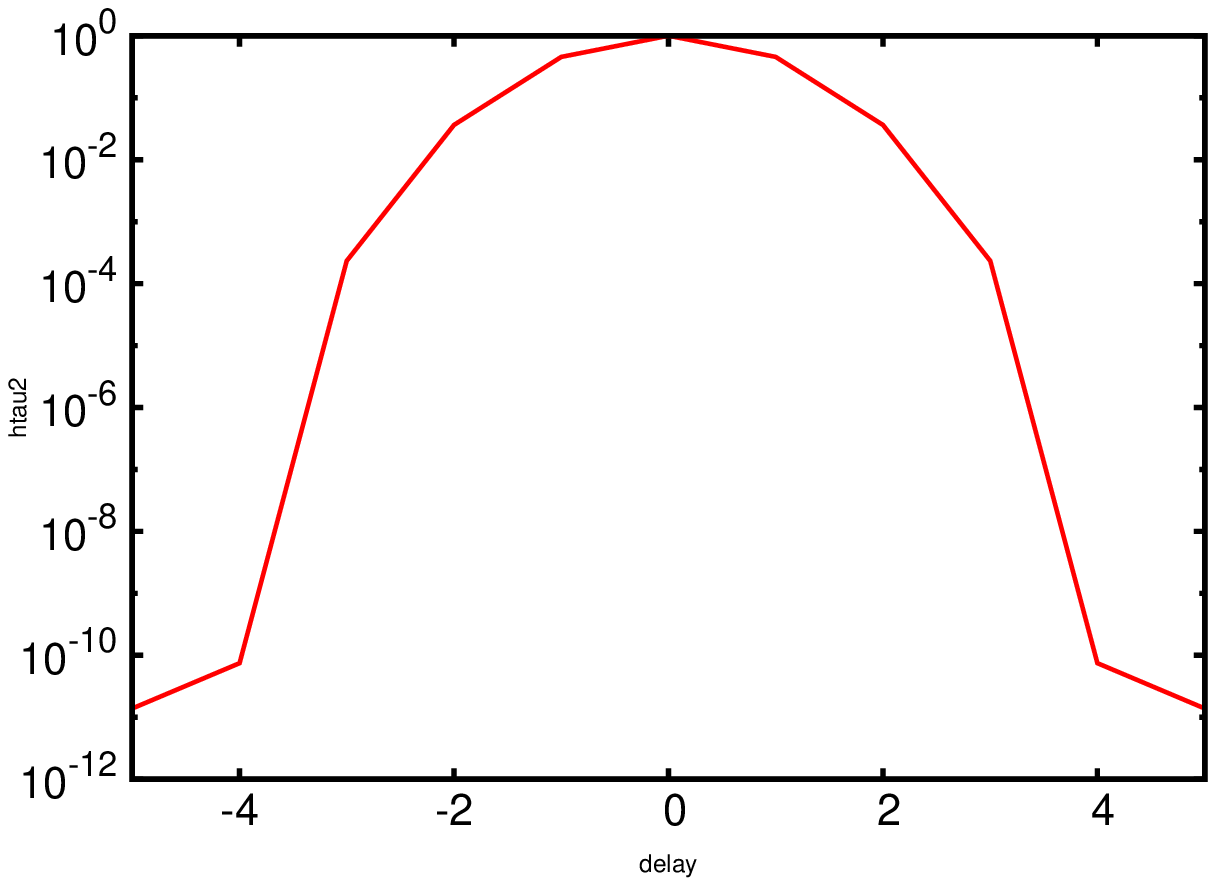}
\caption{The Blackman-Nuttall frequency window  $F(\nu)$ as a function of channel number is 
shown in the left panel. The right panel shows ($\mid\tilde{f}(\tau)\mid^2$)  which is 
the square of the Fourier transform of $F(\nu)$ . This is normalized to unity at the central delay channel.}  
\label{fig:fig3}
\end{center}
\end{figure*}

The right panel of Figure \ref {fig:fig3} show  $\mid\tilde{f}(\tau_m)\mid^2$
as a function of the delay channel number $m$. We see that
$\mid\tilde{f}(\tau_m)\mid^2$ has 
a    very narrow extent  in delay  space, implying that the visibilities 
 $v^f_i(\tau_m)$ in only three adjacent  delay channels are correlated, and  
 $v^f_i(\tau_m)$ are uncorrelated if the delay channel separation is larger
than this. This also allows us to approximate 
$\mid \tilde{f}(\tau_m-\tau_n) \mid^2$ using  a Kronecker delta function
$\approx {\rm B_{bw}^2} \, A_f(0) \, \delta_{m,n}$   where $A_f(0)= \frac{1}{\rm
  B_{bw}^2} \sum_n \mid \tilde{f}(\tau_n) \mid^2$. The convolution in
eq. (\ref{eq:c3}) now gives 
\begin{equation}
\langle \mid v^f_i(\tau_m) \mid^2  \rangle=A_f(0) \, \langle \mid v_i (\tau_m)\mid^2
\rangle \,. 
\label{eq:c4}
\end{equation}
We now generalize this to calculate the correlation for two different values
of $\tau_m$  which gives 
\begin{equation}
\langle v^f_i(\tau_m) v^{f*}_i(\tau_n)   \rangle =A_f(m-n) \, \langle \mid
v_i (\tau_m)\mid^2 \rangle 
\label{eq:c4a} 
\end{equation}
where 
\begin{equation}
A_f(m-n)=\frac{1}{\rm  B_{bw}^2} \sum_a  \tilde{f}(\tau_m-\tau_a) 
\tilde{f}^{*}(\tau_n-\tau_a)
\label{eq:c4b} 
\end{equation}
and $A_f(m-n)=A^{*}_f(n-m)$. We find that $A_f(m)$ has significant values only
for $m=0,1,2,3$ beyond which the values are rather small {\em i.e.} the
visibilities at only the three adjacent delay channels have significant
correlations, and the visibilities are uncorrelated beyond this separation. 
We have used the self-correlation (eq. \ref{eq:c4}) to calculate the power
spectrum estimator later in this subsection,  whereas the general expression for
the correlation (eq. \ref{eq:c4a}) comes in useful for calculating the
variance in a subsequent subsection.

Incorporating the frequency  window function in the  3D TGE introduces an 
additional factor of $A_f(0)$ in the normalization coefficient 
in  eq.  (\ref{eq:a16}).  We now have the final expression for the 3D TGE  
as
\begin{equation}
{\hat P}_g(\tau_m)= \left(\frac{M_g {\rm B_{bw}}\,A_f(0)}{r^2 r^{'}} \right)
^{-1} \, \left( \mid v^f_{cg}(\tau_m) \mid^2 -\sum_i \mid \tilde{w}(\u_g-\u_i)
\mid^2 \mid v^f_i(\tau_m) \mid^2 \right) \,.   
\label{eq:c5}
\end{equation}
As mentioned earlier,  ${\hat P}_g(\tau_m)$  gives an estimate of the power
spectrum  $P({\kpr}_g,{\kp}_m)$   where  ${\kp}_m$ 
and ${\kpr}_g$ are related to $\tau_m$ and $\u_g$ through eqs.   
(\ref{eq:b7}) and (\ref{eq:b8}) respectively.

\subsection{Binning and Variance}
\label{3dvar}
The  estimator ${\hat P}_g(\tau_m)$ presented in eq. (\ref{eq:c5}) provides an 
estimate of  the 3D power spectrum  $P({\kpr}_g,{\kp}_m)$  at an individual  
grid point $\k=({\kpr}_g,{\kp}_m)$ in  the three dimensional $\k$ space. 
Usually one would like to average  the estimated  power spectrum  over a bin
in $\k$ space  in order to increase the signal-to-noise ratio.   In this section we discuss the bin averaged 3D TGE and
obtain formulas for theoretically predicting the expected variance.

We  introduce   the binned 3D TGE which for the  bin labeled   $a$ is  
defined as   
\begin{equation}
{\hat P}_G (a)= \frac{\sum_{gm} w_{gm}  {\hat P}_g(\tau_m)}{\sum_{gm} w_{gm }}
\,
\label{eq:3dvar1}
\end{equation}
where the sum  is over all the $\k=({\kpr}_g,{\kp}_m)$  modes  or equivalently
the grid points ($\u_g$,$\tau_m$) included in the particular bin $a$, and   
$ w_{gm}$ is the weight assigned to  the contribution from any
particular grid point.  Earlier in this paper, in the discussion just
subsequent to eq. (\ref{eq:a15}),  we have introduced the weighing 
scheme $w_{g}=1$ in order to calculate $C_{\ell}$. Here we have adopted the 
same scheme $w_{gm}=1$  for estimating the 3D
power spectrum.  

The expectation value of the binned 3D TGE (eq.~\ref{eq:3dvar1}) 
\begin{equation}
\langle {\hat P}_G (a) \rangle = \bar{P}(\bar{k}_\perp,\bar{k}_\parallel)_a
\label{eq:3dvar1a}
\end{equation}
gives an
estimate of the bin averaged 3D power spectrum 
\begin{equation}
\bar{P}(\bar{k}_\perp,\bar{k}_\parallel)_a= \frac{\sum_{gm} w_{gm}  P({\kpr}_g,{\kp}_m)}{\sum_{gm} w_{gm }} \, 
\label{eq:3dvar2}
\end{equation}
at
\begin{equation}
(\bar{k}_\perp,\bar{k}_\parallel)_a=\Big(\frac{\sum_{gm} w_{gm}
    {\kpm}_g}{\sum_{gm} w_{gm }} ,\frac{\sum_{gm} w_{gm}  {\kp}_m}{\sum_{gm}
    w_{gm }}\,\Big). 
\label{eq:3dvar3}
\end{equation}
where for the particular bin $a$ the two components 
$(\bar{k}_\perp,\bar{k}_\parallel)_a$  refer to the  average  wave numbers 
respectively  perpendicular and parallel to the line of sight.  In this paper
we have considered two different binning schemes which we discuss later in this
sub-section. For the present, we turn our attention to calculate theoretical
predictions for the variance of the binned 3D TGE. 

The variance calculation closely follows the steps outlined in
section \ref{sec:clvar1}, and we  have the final expression 
\begin{equation}  
\sigma^2_{P_G} = \left(\frac{ {\rm B_{bw}}\,A_f(0)}{r^2 r^{'}} \right) ^{-2}
\frac{\sum_{gm,g^{'} m^{'}} \, w_{gm} w_{g^{'}m^{'}}  M_g^{-1} 
  M^{-1}_{g^{'}} \mid \langle v^f_{cg}(\tau_m) v^{f*}_{cg^{'}}(\tau_{m^{'}})
  \rangle \mid^2 }{[\sum_{gm} w_{gm}]^2} \,. 
\label{eq:3dvar4}
\end{equation}
which closely resembles eq. (\ref{eq:var2}) which we have used to calculate
the variance for $C_{\ell}$,  with the difference that we now 
have a 3D grid instead of the 2D grid encountered earlier for $C_{\ell}$. 

It is necessary to model the term 
$\langle v^f_{cg}(\tau_m) v^{f*}_{cg^{'}}(\tau_{m^{'}})
  \rangle $ in eq. (\ref{eq:3dvar4}) to make further progress.  
The correlation at two different $\tau_m$ values can be  expressed  
using eq. (\ref{eq:c4a}) as 
\begin{equation} 
 \langle v^f_{cg}(\tau_m) v^{f*}_{cg^{'}}(\tau_{m^{'}}) \rangle=A_f(m-m^{'})
 \langle v_{cg}(\tau_m) v^{*}_{cg^{'}}(\tau_{m}) \rangle \,.
\label{eq:3dvar5}
\end{equation}
Following eq. (\ref{eq:var2a}), we have decomposed 
the correlation $ \langle v_{cg}(\tau_m) v^{*}_{cg^{'}}(\tau_{m}) \rangle $
in eq. (\ref{eq:3dvar5})  into two parts 
\begin{equation}
\langle v_{cg}(\tau_m) v^{*}_{cg^{'}}(\tau_{m}) \rangle = \langle s_{cg}(\tau_m) s^{*}_{cg^{'}}(\tau_{m}) \rangle + \langle n_{cg}(\tau_m) n^{*}_{cg^{'}}(\tau_{m}) \rangle 
\label{eq:3dvar6}
\end{equation}
corresponding to the signal and the noise respectively. 

We have modeled  the signal correlation in exact analogy with eq. (\ref{eq:var3})
as 
\begin{equation}
\langle s_{cg}(\tau_m) s^{*}_{cg^{'}}(\tau_m) \rangle = \left(\frac{ {\rm
    B_{bw}}}{r^2 r^{'}} \right) \sqrt{M_g M_{g^{'}}} \, 
e^{-\mid \Delta  \u_{g g^{'}} \mid^2/\sigma_1^2}
\bar{P}(\bar{k}_\perp,\bar{k}_\parallel)_a  \,
\label{eq:3dvar7}
\end{equation}
and the noise correlation is similarly  modeled in exact analogy  with
eq. (\ref{eq:var5}) as 
\begin{equation}
\langle n_{cg}(\tau_m) n^{*}_{cg^{'}}(\tau_m) \rangle = (\Delta \nu_c){\rm
  B_{bw}}\sqrt{K_{2gg}K_{2g^{'}g^{'}}}e^{-   \mid \Delta \u_{g g^{'}}
  \mid^2/\sigma_2^2}(2\sigma_n^2)\,.  
\label{eq:3dvar8}
\end{equation}

We have used eqs. (\ref{eq:3dvar8}), (\ref{eq:3dvar7}), (\ref{eq:3dvar6}),
(\ref{eq:3dvar5}) and (\ref{eq:3dvar4}) to calculate the variance of the
binned 3D TGE. In the subsequent analysis  
we have considered two different binning schemes which we now present below. 
\begin{figure*}
\begin{center}
\psfrag{kx}[c][c][1.2][0]{${\bf k}_x$}
\psfrag{ky}[c][c][1.2][0]{${\bf k}_y$}
\psfrag{kpar}[c][c][1.2][0]{$\kp$}
\psfrag{ka}[c][c][1.2][0]{$k_a$}
\psfrag{dka}[c][c][1.2][0]{$\Delta {k}_a$}
\psfrag{kpara}[c][c][1.2][0]{${\kp}_a$}
\psfrag{dkpara}[c][c][1.2][0]{$\Delta {\kp}_a$}
\psfrag{kpera}[c][c][1.][0]{${\kpr}_a$}
\psfrag{dkpera}[r][r][1.][0]{$\Delta {\kpr}_a$}
\hspace*{-1.5cm}
\includegraphics[width=60mm,angle=0]{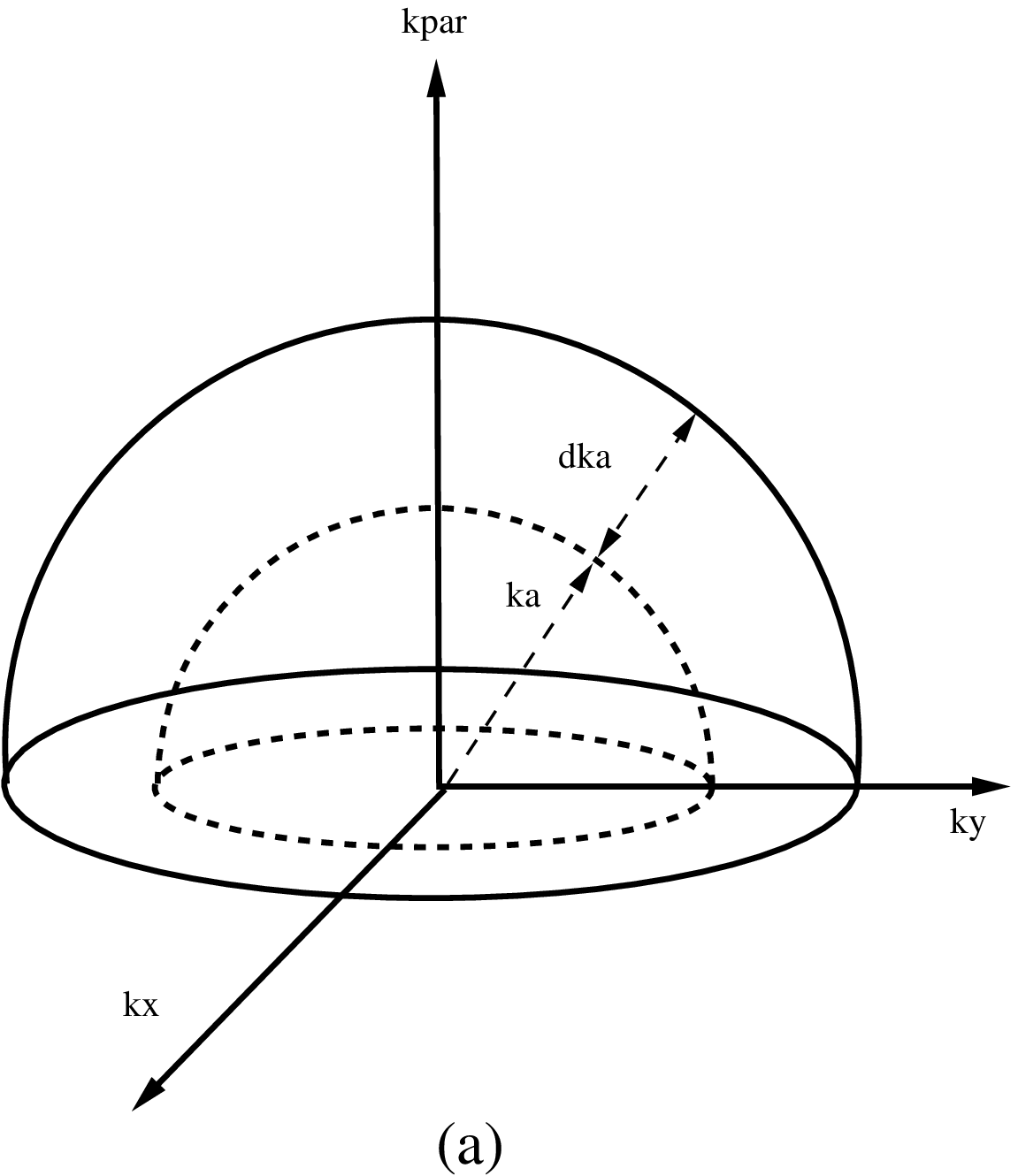}
\hspace*{2cm}
\includegraphics[width=60mm,angle=0]{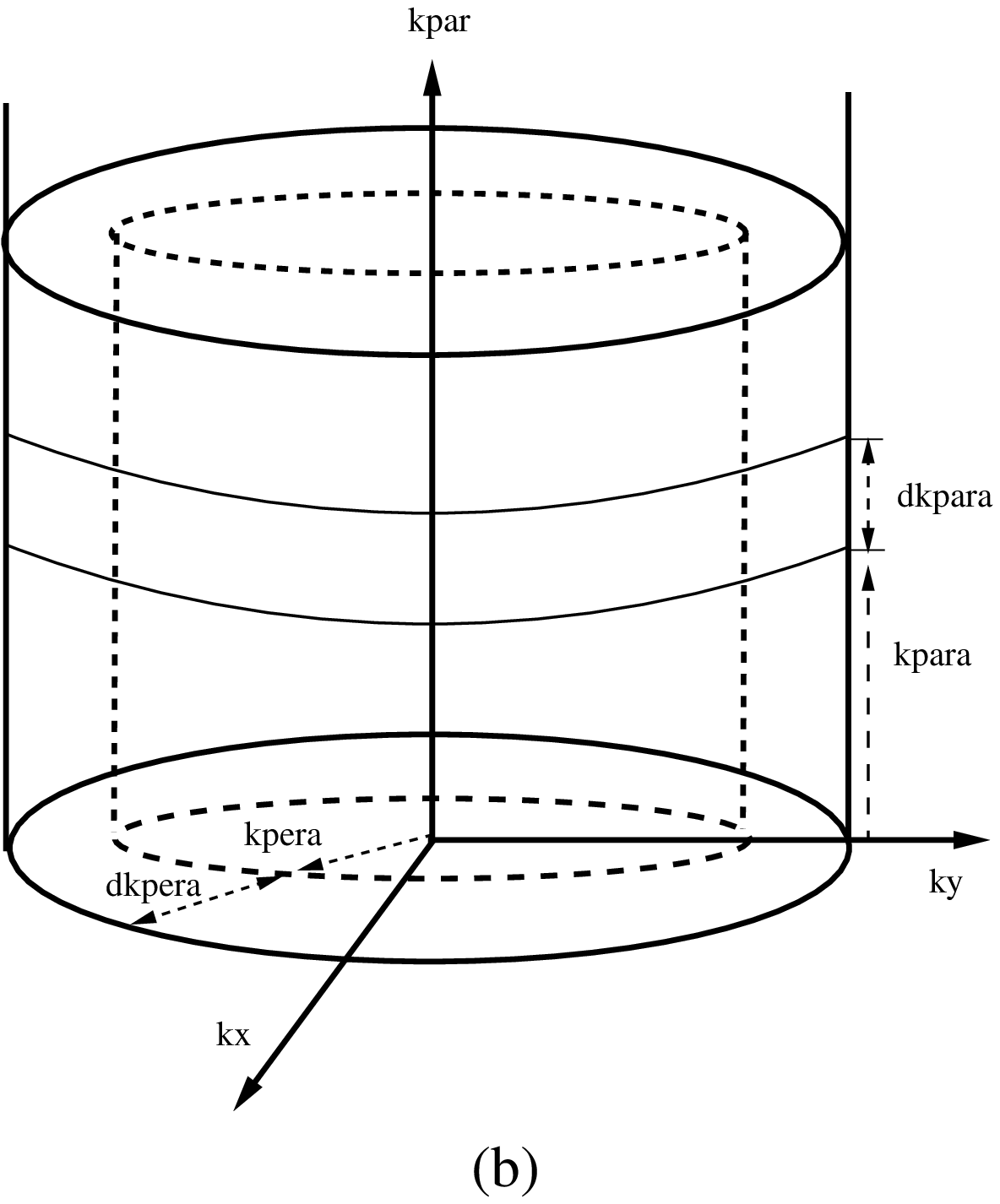}
\caption{This shows a typical bin for respectively calculating the Spherical Power Spectrum 
(left) and the Cylindrical Power Spectrum (right).}
\label{fig:bin}
\end{center}
\end{figure*}

\subsubsection{1D Spherical Power Spectrum}
The bins here are spherical shells of thickness $\Delta k_a$ as shown in the 
left panel of Figure \ref{fig:bin}, the shell thickness will in general vary
from bin to bin. The  Spherical Power Spectrum $\bar{P}(\bar{k}_a)$
is obtained by averaging the power spectrum $P(\k)$  over all the different
$\k$ modes which lie within the spherical shell corresponding to bin $a$ 
shown in the  left panel of Figure \ref{fig:bin}.  The  binning here
essentially averages out any anisotropy in the power spectrum, and yields the
bin averaged power spectrum as a function of the 1D bin averaged wave number
$\bar{k}_a$. While  
we use  eq. (\ref{eq:3dvar1}) to calculate the bin averaged power spectrum
$\bar{P}(\bar{k}_a)$, 
we have calculated the value of  $\bar{k}_a$ using 
\begin{equation}
 \bar{k}_a=\frac{\sum_{gm} w_{gm}    \sqrt{ {\kpm}_g^2 + {\kp}_m^2}}
     {\sum_{gm} w_{gm }} \,. 
\end{equation}

\subsubsection{2D Cylindrical Power Spectrum}
\label{sec:cylnbin}
Each bins here is, as shown in the  right panel of Figure \ref{fig:bin},   an
annulus of width $\Delta {\kpm}_a$ in the $\kpr \equiv (k_x,k_y)$ plane and it
subtends a thickness $\Delta {\kp}_a$ along the third direction $\kp$.  The
values of  $\Delta {\kpm}_a$ and $\Delta {\kp}_a$ will, in general, vary from bin
to bin. The bins here correspond to sections of a hollow cylinder, and the
resulting bin averaged power spectrum
$\bar{P}(\bar{k}_\perp,\bar{k}_\parallel)_a$ is referred to as the
Cylindrical Power Spectrum which   is defined on a 2D space 
$(\bar{k}_\perp,\bar{k}_\parallel)_a$  whose two components   refer to the
average  wave numbers  respectively  perpendicular and parallel to the line of
sight.  The binning of $P(\k)$ here does not assume that the signal is
statistically isotropic in the 3D space {\em i.e.} independent of the direction
of  $\k$. However, the signal  is assumed to be statistically
isotropic in the plane of the sky, and the binning in 
$\kpr$ is exactly identical to the binning that we
have used earlier for $C_{\ell}$. This distinction between $\kpm$ and $\kp$ is
useful to quantify the effect of redshift space distortion
\citep{BNS01,BS01,BA2004,barkana05,mao12,majumdar13,jensen16} and also to distinguish the foregrounds
from the HI signal \citep{morales04}.  We have used  eq. (\ref{eq:3dvar1}) and
eq. (\ref{eq:3dvar3})   to calculate
$\bar{P}(\bar{k}_\perp,\bar{k}_\parallel)_a$ and
$(\bar{k}_\perp,\bar{k}_\parallel)_a$ respectively.

\section{Simulation}
\label{simu}
In this section we discuss the  simulations that we have used 
to validate the 3D power spectrum estimator (eq. \ref{eq:c5}). We start with
 an input model 3D power spectrum $P^M(k)$ of 
redshifted HI 21-cm brightness temperature fluctuations. 
The aim here is to test  how well the estimator is able to recover the
input model. For this purpose  the exact form of the input model power
spectrum  need not mimic the  expected cosmological HI signal, and we have
used a simple power law 
\begin{equation}
P^M(k)=\left(\frac{k}{k_0} \right)^n
\label{eq:pk1}
\end{equation}
which is arbitrarily normalized to unity at $k=k_0$, and has a power law index
$n$. In our analysis we have considered $n=-3$ and $-2$, and set 
$k_0=1 \, {\rm  Mpc}^{-1}$.  The quantity $\Delta_k^2=(2 \pi^2)^{-1} k^3 P(k)$
provides an estimate of the mean-square brightness  temperature 
fluctuations expected at different length-scales (or equivalently 
 wave numbers  $k$). We see that for $n=-3$ we have a constant 
 $\Delta_k^2= (2 \pi^2)^{-1}\,  {\rm K^2}$  across all  length-scales, whereas we have 
$\Delta_k^2= (2 \pi^2)^{-1} (k/1 \, {\rm Mpc}^{-1}) \, {\rm K^2}$ which increases
linearly with $k$ for $n=-2$. Note that we have used an isotropic  
input model where the power spectrum does not  depend on the direction of
$\k$ {\em i.e.}  $(P(\k) \equiv P(k))$ and  the 1D Spherical binning
and the 2D Cylindrical binning are expected to recover the same results.   

The simulations were carried out using a $N^3$ cubic grid of spacing $L$
covering a comoving volume $V$. We use the model power spectrum
(eq. \ref{eq:pk1}) to generate the Fourier components of the brightness
temperature fluctuations corresponding to this grid 
\begin{equation}
\Delta {\tilde T}(\mathbf{k}) = \sqrt{\frac{V P^M(k)}{2}} [a(\mathbf{k})+\mathit{i}
  b(\mathbf{k})] \,,
\label{eq:pk2}
\end{equation}
here $a(\mathbf{k})$ and $b(\mathbf{k})$ are two real valued independent
Gaussian random variable of unit variance.  The Fourier transform of  $\Delta
T(\mathbf{k})$ yields a single  realization of the brightness temperature
fluctuations $\delta T(\x)$ on the simulation grid. These fluctuations are, by
construction, a Gaussian random field with power spectrum $P^M(k)$. We
generate different statistically independent realizations of  $\delta T(\x)$
by using different sets of random variables  $a(\mathbf{k})$ and
$b(\mathbf{k})$ in eq. (\ref{eq:pk2}). 

The intention here is to simulate $150 \, {\rm MHz}$ GMRT observations with 
$N_c=256$ frequency channels of width $ (\Delta \nu_c)=62.5 \, {\rm kHz}$
covering a bandwidth of ${\rm  B_{bw}}=16 \, {\rm MHz}$. This corresponds to
HI at redshift  $z=8.47$ with  a comoving distance of 
$r = 9.28 \, {\rm  Gpc}$  and $r^{'}=\mid dr/d \nu \mid= 17.16 \,  {\rm Mpc
  \,  MHz}^{-1}$.  We have chosen the grid spacing  $L=1.073 \, {\rm Mpc}$  so
that it exactly matches the   channel width  $L =    r_{\nu}' \times (\Delta
\nu_c)$.  We have considered a $N^3=[2048]^3$ grid which corresponds to a
comoving volume of $[2197.5 \, {\rm Mpc}]^3$. The simulation volume is aligned
with the $z$ axis along the line of sight, and  the two transverse directions were
converted to angles relative to the box center
$(\theta_x,\theta_y)=(x/r,y/r)$.   The transverse extent of the 
simulation box covers an angular extent which is $\sim 5$ times the GMRT
$\theta_{FWHM}$.  The simulation volume corresponds to a frequency width $\sim
8 \times 16 \, {\rm MHz}$ along the line of sight. We have cut the box into $8$
equal segments along the line of sight to produce $8$ independent realizations
each subtending $16 \, {\rm MHz}$ along the line of sight. The grid index,
measured from the further boundary and increasing towards to observer along
the line of sight was directly converted to channel number $\nu_a$ with
$a=0,1,2,...,N_c-1$.  This procedure provides us with $\delta T(\th,\nu_a)$
the brightness temperature fluctuation on the sky at different frequency
channels $\nu_a$. 

We have considered $8$ hours of  GMRT observations with $16 \, {\rm s}$
integration time targeted on an arbitrarily selected  field  located at
RA=$10{\rm h} \, 46{\rm  m} \, 00{\rm s}$ and DEC=$59^{\circ} \,  00^{'} \,
59^{''}$.  Visibilities were calculated for the simulated baselines
corresponding to this observation, for which the $uv$ coverage is similar to the 
Figure 5 of Paper I. The signal contribution to the visibilities
$\S(\u,\nu_a)$  was calculated by taking the Fourier transform of the product
$\left(\frac{\partial B}{\partial T}\right)  \times 
{\mathcal A}(\th,\nu_a) \times \delta T(\th,\nu_a) $ as given by eq. (\ref{eq:b2a}).  
The simulations incorporate the fact that   the baseline corresponding to a
fixed antenna separation $\u_i={\bf d}_i/\lambda$, the antenna beam pattern
${\mathcal A}(\th,\nu_a)$    and the factor $\left( \frac{\partial B}{\partial
  T} \right)_{\nu_a}$ all vary with the frequency $\nu_a$ in
eq. (\ref{eq:b2a}). We have $\sigma_n= 1.45\, {\rm Jy}$ corresponding to a single
polarization, with $\Delta t=16 \, {\rm s}$ and $ (\Delta \nu_c)=62.5 \,
{\rm kHz}$. However, it is possible to reduce noise level by averaging independent data set
observed at different time. Here, we consider a situation where we average $9$ independent data sets 
to reduce the noise level by a factor of  $3$ to $\sigma_n= 0.48\, {\rm Jy}$. We have carried out  
the  simulations for 
two different cases, (i) no noise ($\sigma_n= 0\, {\rm Jy}$) and (ii) $\sigma_n= 0.48\, {\rm Jy}$.  We have  
carried out  $16$ independent  realization of the simulated visibilities to estimate the mean 
power spectrum and its statistical fluctuation (or standard deviation $\sige$)
presented in the next section.

\section{Results}
\label{result}
 The left panels of Figures \ref{fig:fig5} and \ref{fig:fig6} show
 $\Delta_k^2=(2 \pi^2)^{-1} k^3 P(k)$ for the spherically-averaged
 power spectrum for the power law index values $n=-3$ and $-2$
 respectively.    The results are shown for the three
 values $f=10 ,2$ and $0.6$ to demonstrate the effect of varying the tapering.
 The simulations here do not include the system noise contribution.
 For both $n=-3$ and $-2$, and for all the values of $f$ we find that
 $\Delta_k^2$ estimated using the 3D TGE is within the
 $1-\sigma_{P_G}$ error bars of the model prediction for the entire
 $k$ range considered here. The right panels of Figures \ref{fig:fig5}
 and \ref{fig:fig6} show the corresponding fractional deviations
 $(P(k)-P^M(k)) /P^M(k)$. For comparison, the relative statistical
 fluctuations, $\sigma_{P_G}/P^M(k)$ are also shown by shaded regions
 for different values of $f$. We find that for both cases $n=-3$ and
 $-2$, the fractional deviation is less than $4\%$ at $k > 0.2\,\rm
 {Mpc}^{-1}$. The fractional deviation increases as we go to lower
 $k$ bins. The fractional deviation also increases if the value of
 $f$ is reduced. The maximum fractional deviation has a value $\sim
 40\%$ and $\sim 20\%$ at the smallest $k$ bin for $n=-3$ and $-2$
 respectively. We find that the fractional deviation is within
 $\sigma_{P_G}/P^M(k)$ for $k \le 0.3\,\rm {Mpc}^{-1}$ and is slightly
 larger than $\sigma_{P_G}/P^M(k)$ for $k \ge 0.3\,\rm {Mpc}^{-1}$. Our results 
indicate  that the 3D TGE is able to recover the model power spectrum to a
reasonably good  level of accuracy ($ \le  20 \%)$ at  the $k$ modes 
$k \ge  0.1\,\rm {Mpc}^{-1}$. The fractional error at the smaller $k$ bins 
increases as the tapering is increased ($f$ is reduced). It may be noted that 
a similar behaviour was also found for $C_{\ell}$ (Figure \ref{fig:fig1}). 
As mentioned earlier,  we attribute this discrepancy to the variation of 
signal amplitude within the width of the convolving window 
$\tilde{w}(\u_g-\u_i)$. This explanation is further substantiated by the 
fact that 
the fractional deviation is found to be larger for $n=-3$ where the power
spectrum is steeper compared to $n=-2$.

\begin{figure*}
\begin{center}
\psfrag{k}[b][b][1.3][0]{$k$ [$\rm {Mpc}^{-1}$]}
\psfrag{k3pk}[c][c][1.3][0]{$\Delta_k^2\,[\rm K^2]$}
\psfrag{model}[r][r][1][0]{Model}
\psfrag{tap10}[r][r][1][0]{f=10}
\psfrag{tap2}[r][r][1][0]{f=2}
\psfrag{tap0.6}[r][r][1][0]{f=0.6}
\psfrag{diffpk}[b][b][1.1][0]{$(P(k)-P^M(k)) /P^M(k),\sigma_{P_G}/P^M(k)$}
\includegraphics[width=80mm,angle=0]{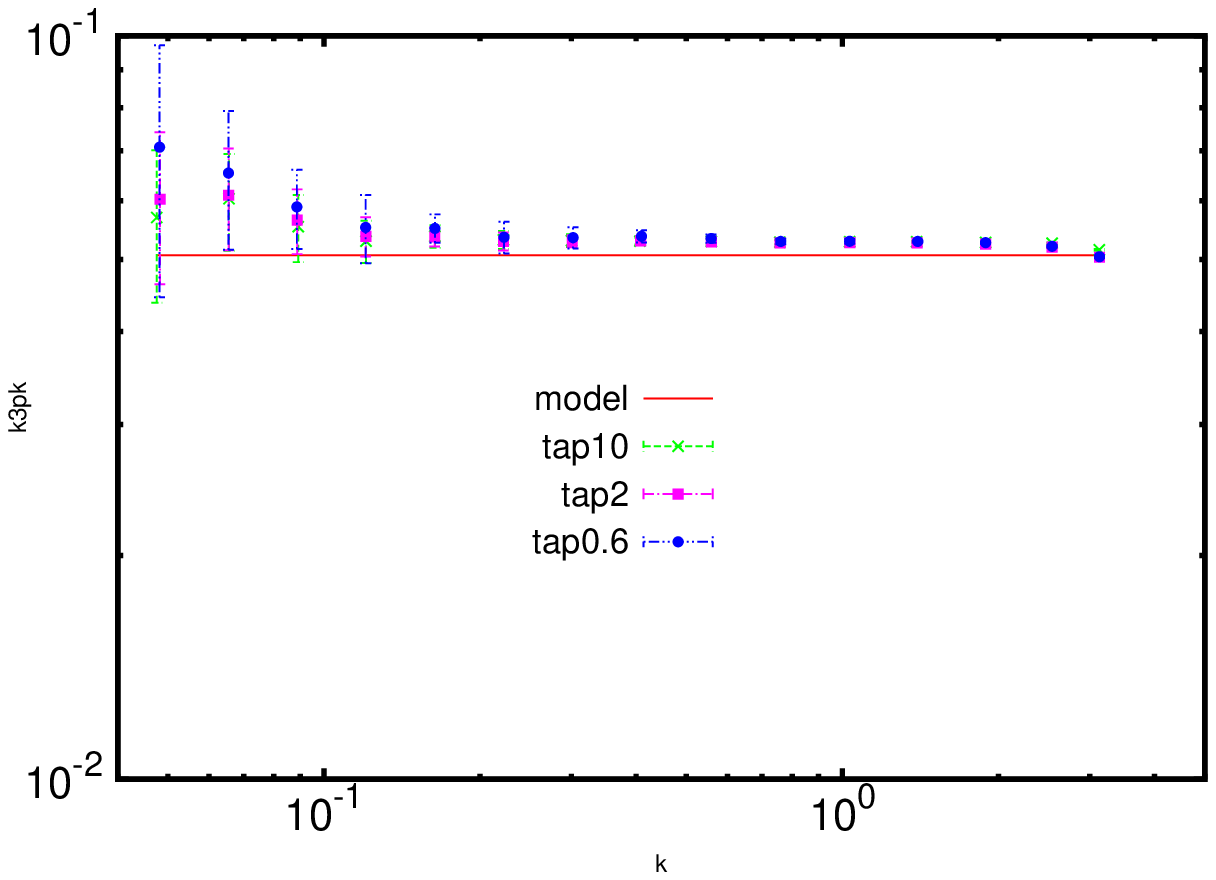}
\includegraphics[width=80mm,angle=0]{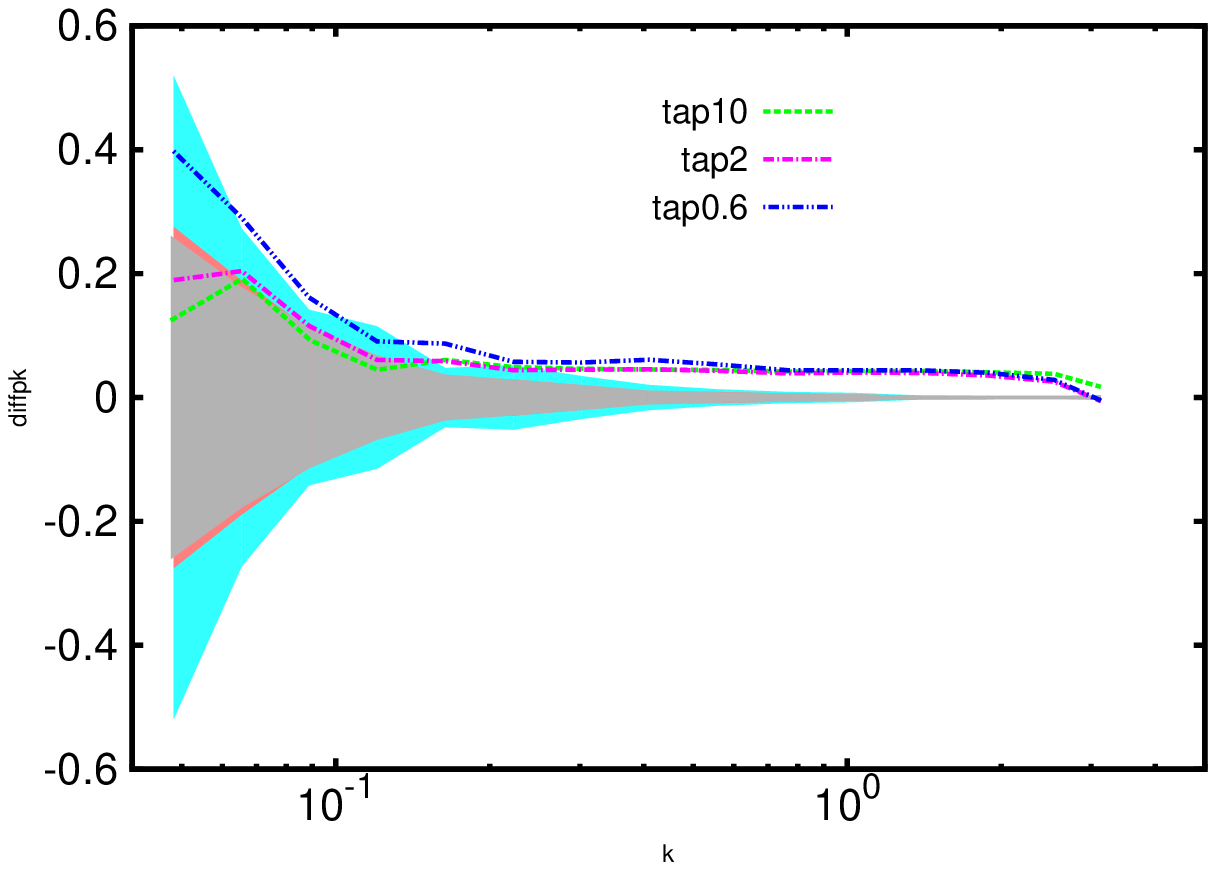}
\caption{The left panel shows the dimensionless power spectrum $\Delta_k^2$ for different values of $f$. The values obtained using the 3D TGE are compared with model power spectrum for $n=-3$ and $\sigma_n= 0 $. The 1-$\sigma_{P_G}$ error bars have been estimated using 16 different realizations of the simulated visibilities. The right panel shows the fractional deviation of estimated power spectrum, $(P(k)-P^M(k)) /P^M(k)$ relative to the input model  $P^M(k)$ for different values of $f$. The relative statistical fluctuations $\sigma_{P_G}/P^M(k)$ are also shown by shaded regions.}
\label{fig:fig5}
\end{center}
\end{figure*}
  
\begin{figure*}
\begin{center}
\psfrag{k}[b][b][1.3][0]{$k$ [$\rm {Mpc}^{-1}$]}
\psfrag{k3pk}[b][c][1.3][0]{$\Delta_k^2\,[\rm K^2]$}
\psfrag{model}[r][r][1][0]{Model}
\psfrag{tap10}[r][r][1][0]{f=10}
\psfrag{tap2}[r][r][1][0]{f=2}
\psfrag{tap0.6}[r][r][1][0]{f=0.6}
\psfrag{diffpk}[b][b][1.1][0]{$(P(k)-P^M(k)) /P^M(k),\sigma_{P_G}/P^M(k)$}
\includegraphics[width=80mm,angle=0]{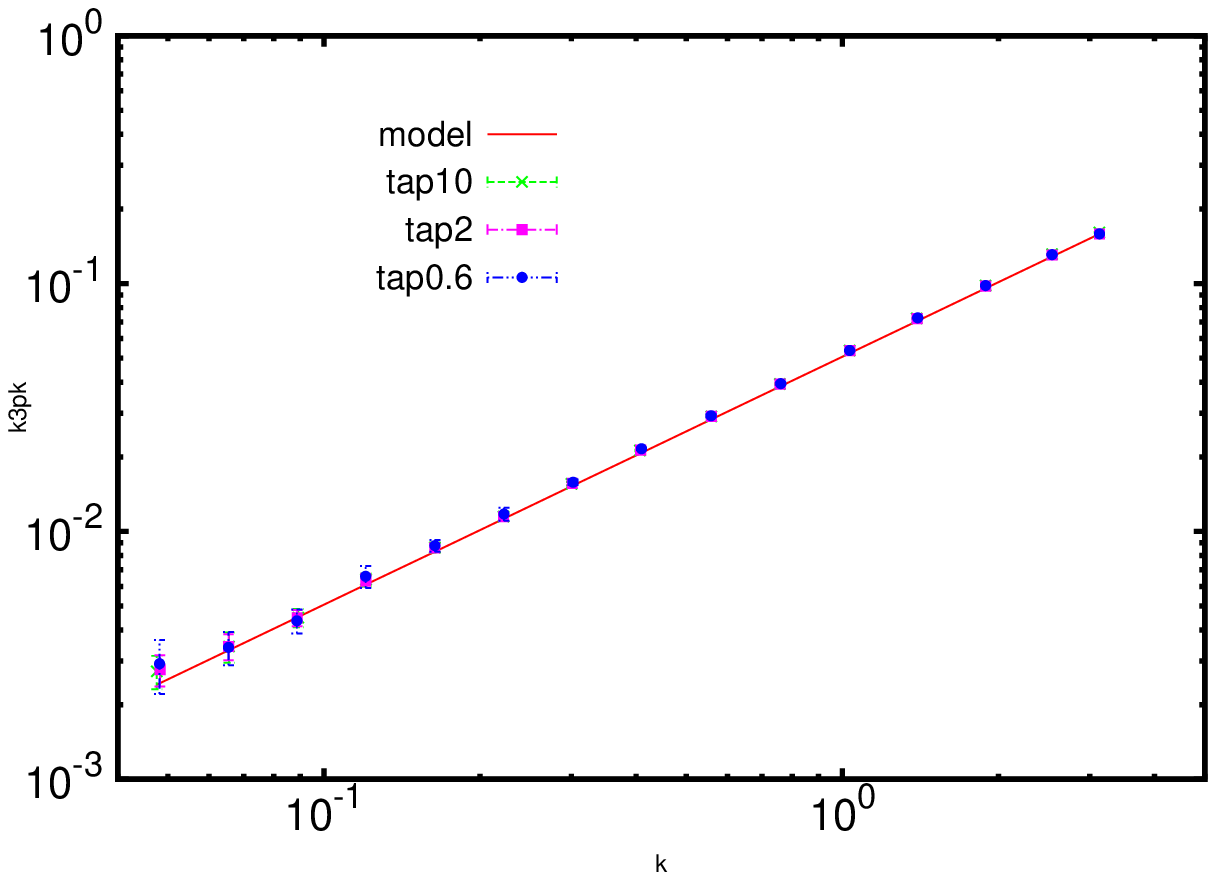}
\includegraphics[width=80mm,angle=0]{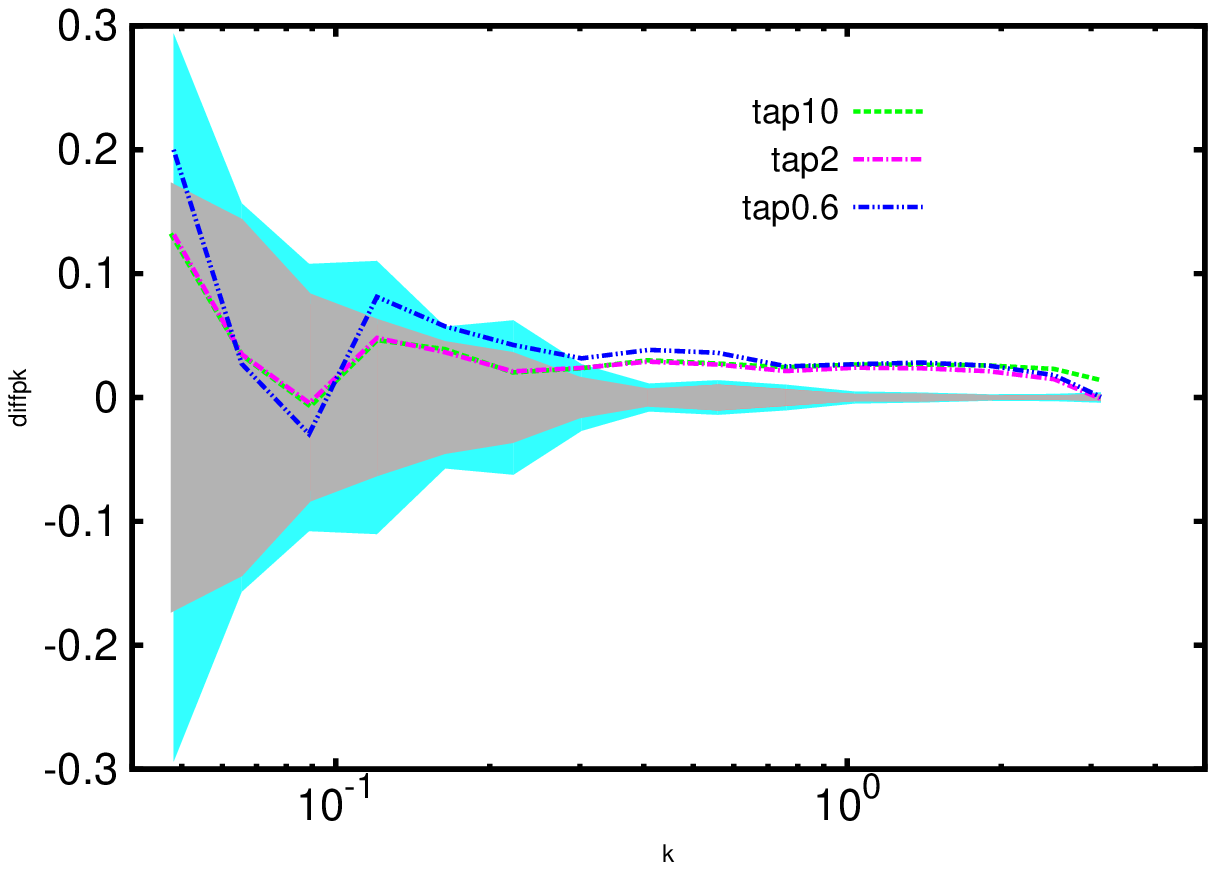}
\caption{Same as Figure~\ref{fig:fig5}, but with $n=-2$.}
\label{fig:fig6}
\end{center}
\end{figure*}

\begin{figure*}
\begin{center}
\psfrag{k}[b][b][1.4][0]{$k$ [$\rm {Mpc}^{-1}$]}
\psfrag{k3pk}[b][c][1.3][0]{$\Delta_k^2\,[\rm K^2]$}
\psfrag{model}[r][r][1][0]{Model}
\psfrag{nonse}[r][r][1][0]{No Noise}
\psfrag{nse}[r][r][1][0]{$\sigma_n=0.48\,{\rm Jy}$}
\psfrag{tap0.6}[r][r][1][0]{f=0.6}
\includegraphics[width=80mm,angle=0]{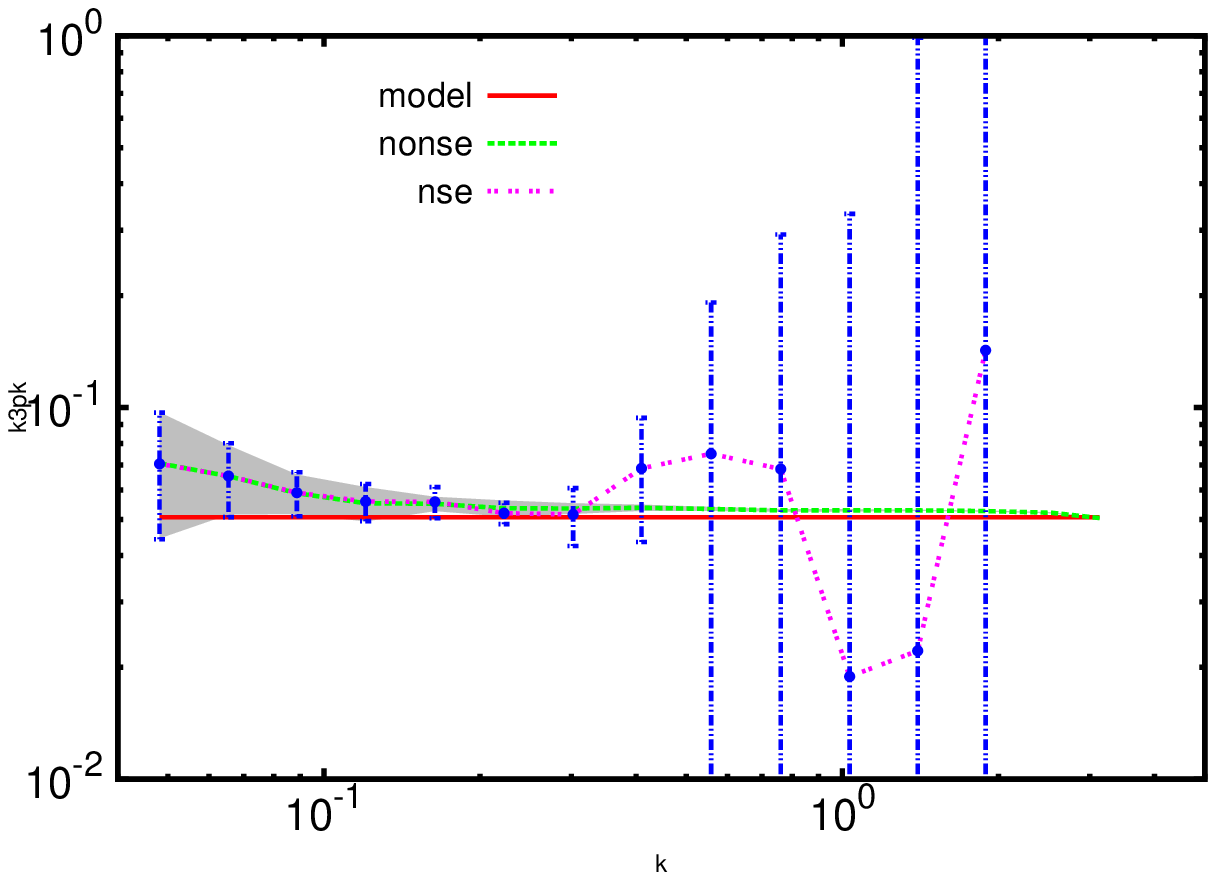}
\includegraphics[width=80mm,angle=0]{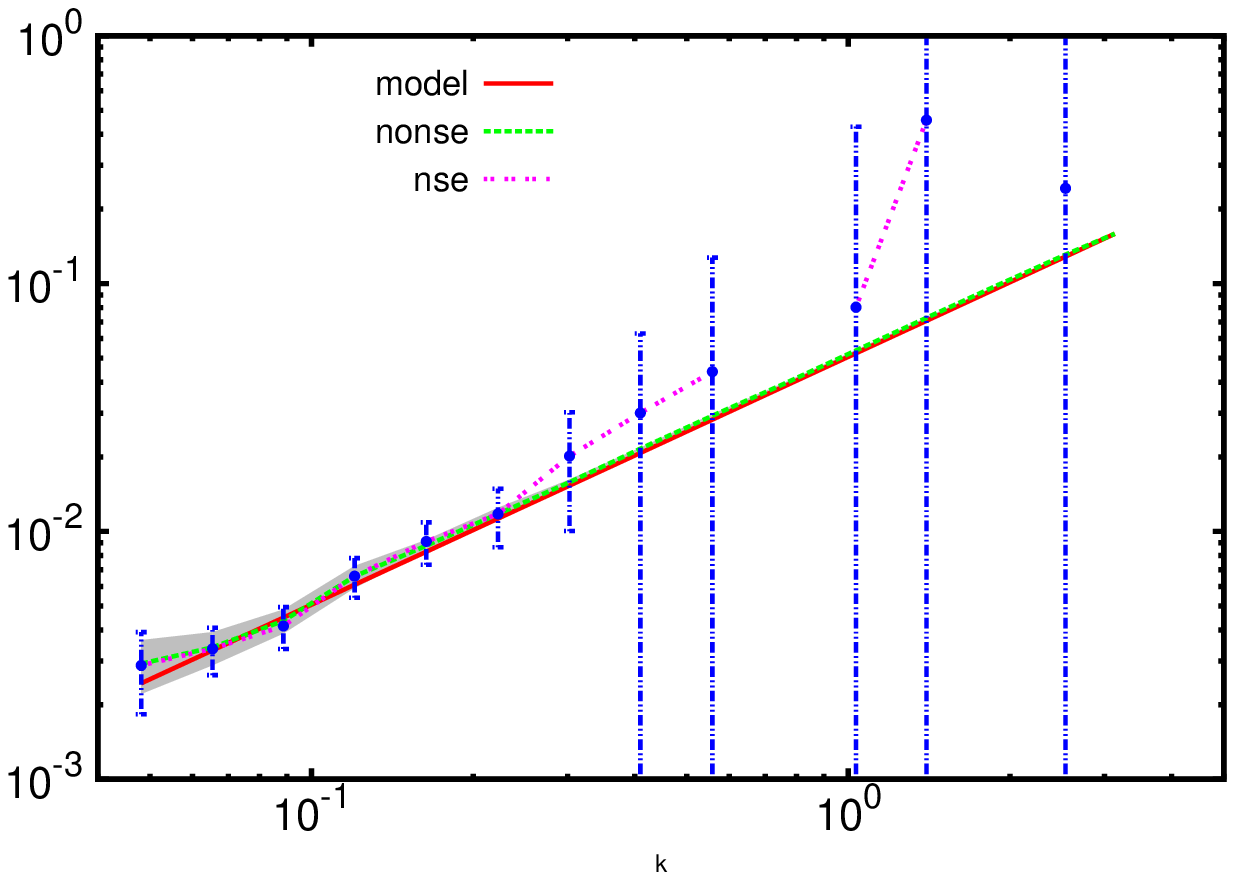}
\caption{The recovered dimensionless power spectrum $\Delta_k^2$ for
  $n=-3$ (left) and $n=-2$ (right), with and without noise for a fixed
  value $f=0.6$.  The statistical error (1-$\sige$) with (without)
  noise is shown with error bars (shaded region). Note that, the estimated
$\Delta_k^2$ has negative values at some of the $k$ values in the range where noise dominates the signal. These data points have not been displayed here.}
\label{fig:fig7}
\end{center}
\end{figure*}

The results until now have not considered the effect of system
noise. We now study how well the 3D TGE is able to recover the input
power spectrum in the presence of system noise. The left and right
panels of Figure \ref{fig:fig7} show the estimated $\Delta_k^2$ for
$n=-3$ and $-2$ respectively for the fixed value $f=0.6$. For
comparison, we also show the estimated $\Delta_k^2$ with $\sigma_n= 0
$. The statistical fluctuations with (without) noise are shown as error
bars (shaded region). We see that the error is dominated by the cosmic
variance at lower values of $k$ ($k < 0.2\,{\rm {Mpc}^{-1}}$) and the
system noise dominates at larger values of $k$. The statistical error
exceeds the model power spectrum at large $k$ and a statistically
significant estimate of the power spectrum is not possible in this $k$
range. We are able to recover the model power spectrum quite
accurately at low $k$ where $\sigma_{P_G} \le P^M(k)$.

We now investigate how well the analytic prediction
(eq. \ref{eq:3dvar4}) for $\sige$ compares with the values obtained
from the simulations (Figure \ref{fig:fig8} ) for different values of
$f$.  The number of grid points in each $k$ bin increase with the
value of $k$, and the computation also increases with increasing
$k$.  We have restricted the $k$ range to $(k < 0.4\,{\rm{Mpc}^{-1}})$
in order to keep the computational requirements within manageable
limits. In the left panel we consider the situation where there is no
system noise. Here, the statistical fluctuations correspond to
the cosmic variance.  We see that the analytic predictions are in
reasonably good agreement with the simulation for both the values of
$f$. We find that the cosmic variance does not change if the value of
$f$ is changed from $2$ to $10$. As expected, the cosmic variance
increases as the sky tapering is increased.  The right panel shows the
statistical fluctuations with and without noise for the fixed value
$f=0.6$. The statistical fluctuations are dominated by the cosmic
variance at small values of $k$ $(k<0.2\,{\rm {Mpc}^{-1}})$, and the
system noise dominates at large $k$.  As mentioned earlier, the
statistical fluctuations are well modeled by the analytic predictions
in the cosmic variance dominated regime. We find that our analytic
prediction somewhat overestimates $\sige$ in the noise dominated
region.  This overestimate possibly originates from the noise
modelling in eq. (\ref{eq:3dvar4}), we plan to investigate this in
future work.

\begin{figure*}
\begin{center}
\psfrag{k}[b][b][1.3][0]{$k$ [$\rm {Mpc}^{-1}$]}
\psfrag{sigk}[b][c][1.3][0]{$k^{3}\sigma_{P_G}/2\pi^2\,[\rm K^2]$}
\psfrag{sim0.6}[r][r][1][0]{Simulation}
\psfrag{ana0.6}[r][r][1][0]{Analytic}
\psfrag{tap2}[r][r][1][0]{f=2, Simulation}
\psfrag{var2}[r][r][1][0]{Analytic}
\psfrag{tap0.6}[r][r][1][0]{f=0.6, Simulation}
\psfrag{var0.6}[r][r][1][0]{Analytic}
\psfrag{f0.6}[r][r][1][0]{f=0.6}
\includegraphics[width=80mm,angle=0]{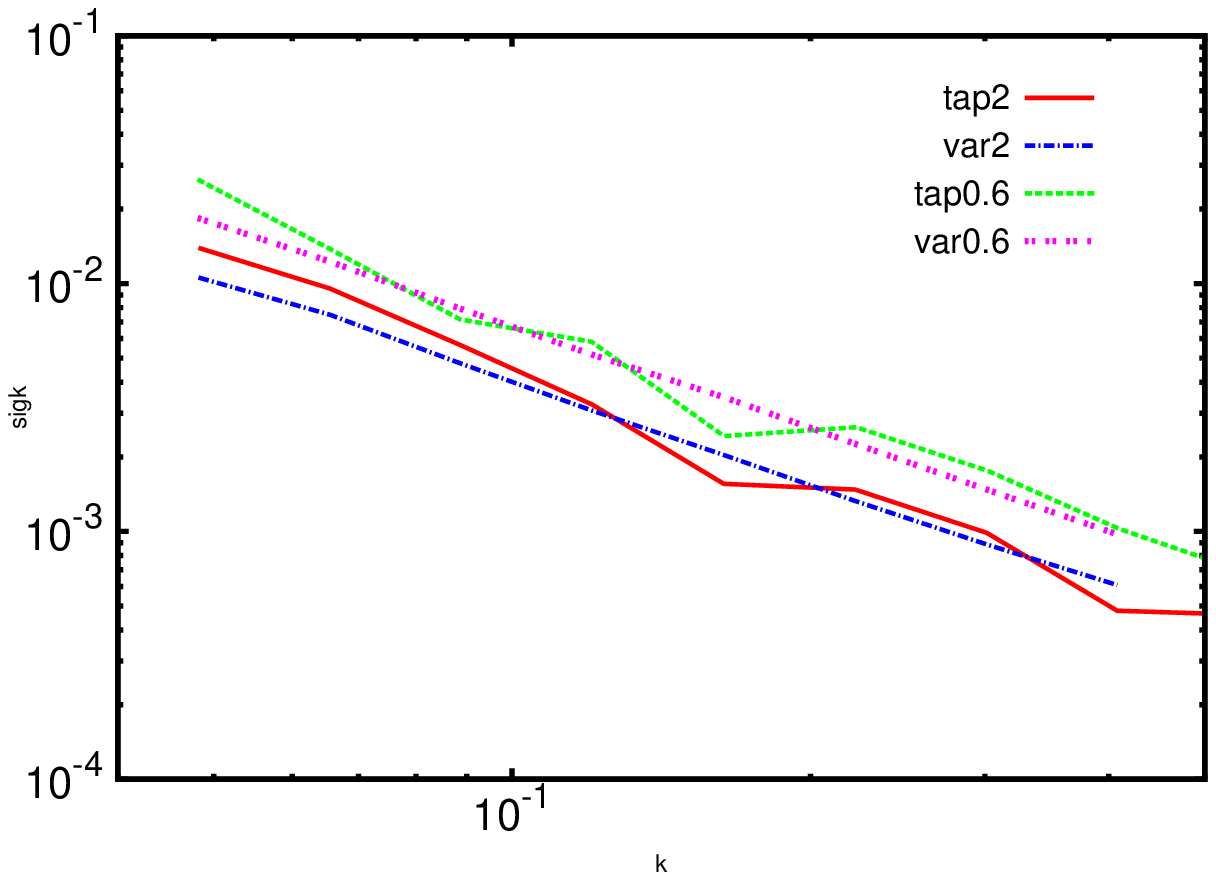}
\includegraphics[width=80mm,angle=0]{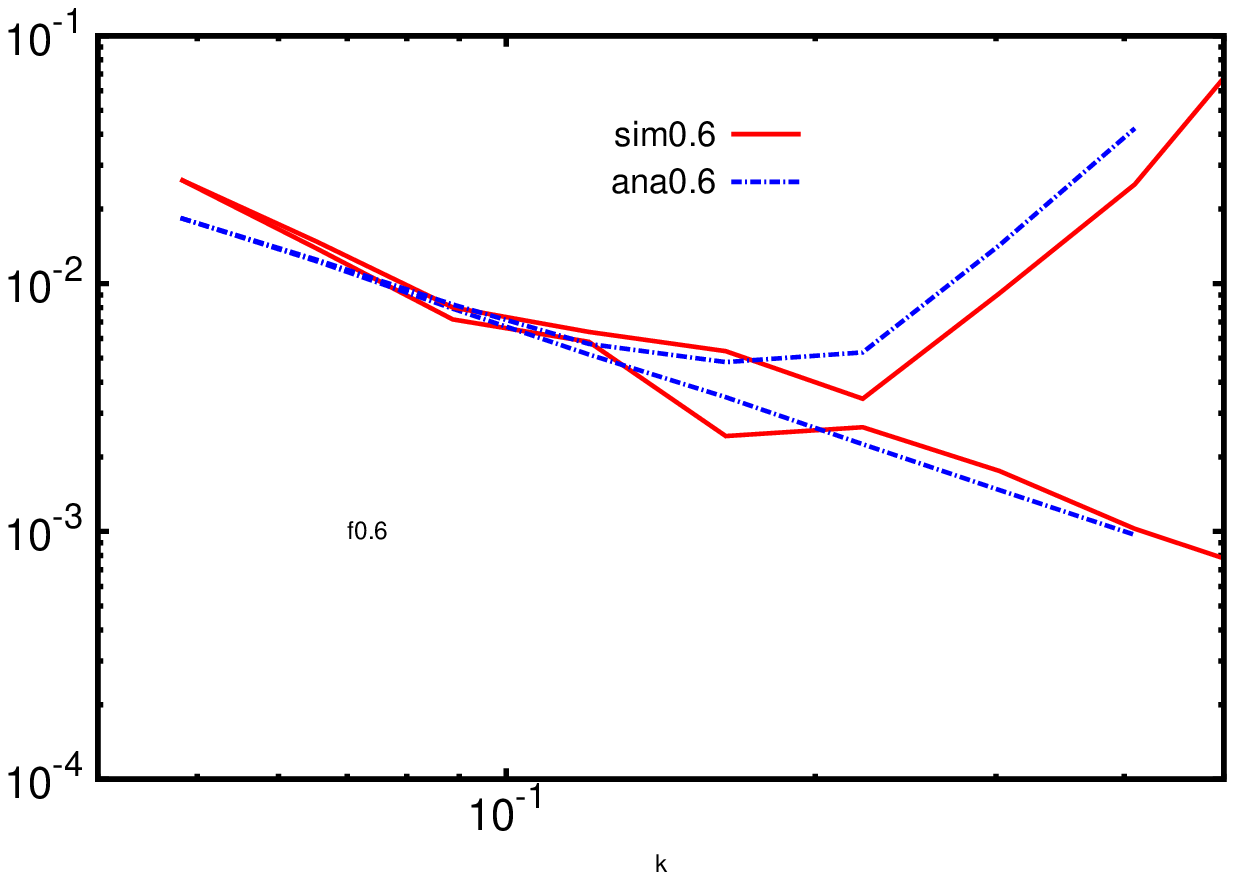}
\caption{The left panel shows a comparison of the analytic prediction
  for the statistical fluctuations of the power spectrum
  (eq. \ref{eq:3dvar4}) with the simulation for two different values
  of $f$, $n=-3$ and no system noise. The right panel shows the same
  comparison with (upper two curves) and without (lower two curves) noise for a fixed value $f=0.6$.}
\label{fig:fig8}
\end{center}
\end{figure*}

Till now we have discussed the results for the 1D Spherical Power
Spectrum, we now present the results for the 2D Cylindrical Power
Spectrum. We use $15$ equally spaced logarithmic bin in both $k_\perp$
and $k_\parallel$ direction to estimate the 2D Cylindrical Power
Spectrum. Figure \ref{fig:fig9} shows the 2D Cylindrical Power
Spectrum $P(k_\perp,k_\parallel)$ using 3D TGE. The left panel shows
the input model for $n=-3$. The middle and right panel respectively
show the estimated power spectrum with $f=0.6$ for situations where
the system noise is not included and included in the simulated
visibilities. The left and middle panels appear almost indistinct
indicating that the 3D TGE is able to recover the input model power
spectrum accurately across the entire $(k_\perp,k_\parallel)$ range.
We find that we are able to recover the model power spectrum in the
limited range $k_\perp \lsim \, 0.5\, \rm {Mpc}^{-1}$ and $k_\parallel
\lsim \, 0.5\, \rm {Mpc}^{-1}$ in presence of system noise.
  Figure \ref{fig:fig9a} shows the fractional deviation
  $(P^M(k_\perp,k_\parallel)-P(k_\perp,k_\parallel))/P(k_\perp,k_\parallel)$
 for $f=0.6$,  here the left and right panels show the results 
without and with system noise  respectively. From the left panel
 we see that the fractional 
  deviation is less than $14\%$ for the the entire $\mathbf{k}$ range
  when the system noise is not included in the simulation. We find that 
it is not possible to reliably recover the power spectrum at large 
$\mathbf{k}$ when the system noise is included. In the right panel 
we have only shown the fractional deviation where it is within $30 \%$, 
the values exceed $100 \%$ at large $\mathbf{k}$ where the values have not been shown.

\begin{figure*}
\begin{center}
\psfrag{kpara}[h][h][1.3][0]{$\kp [\rm {Mpc}^{-1}]$}
\psfrag{kper}[b][b][1.3][0]{$\kpm [\rm {Mpc}^{-1}]$}
\psfrag{model}[r][r][1][0]{}
\psfrag{k2mpc3}[t][c][1.2][0]{log($P(k_\perp,k_\parallel)$)}
\hspace*{-3.0cm}
\includegraphics[width=80mm,height=60mm,angle=0]{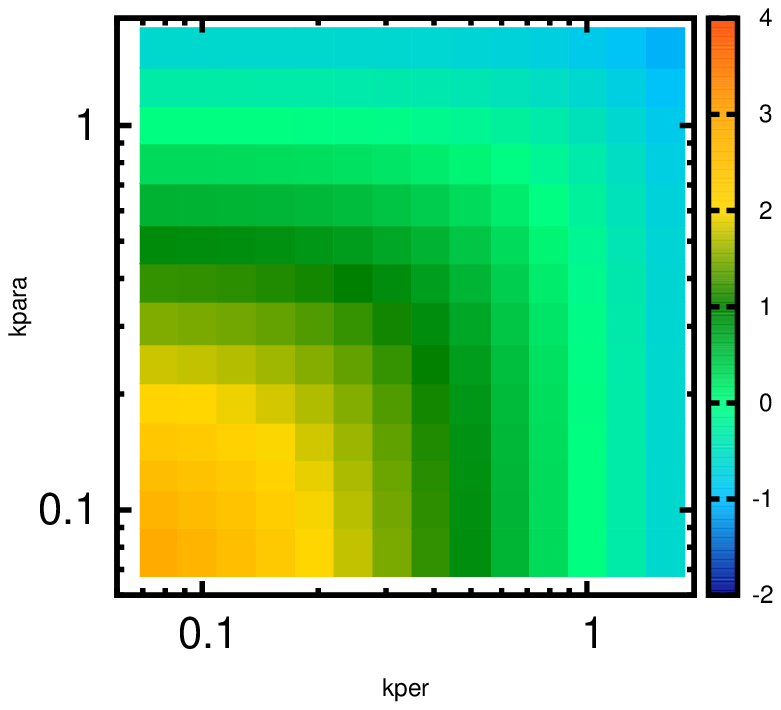}
\hspace*{-2.5cm}
\includegraphics[width=80mm,height=60mm,angle=0]{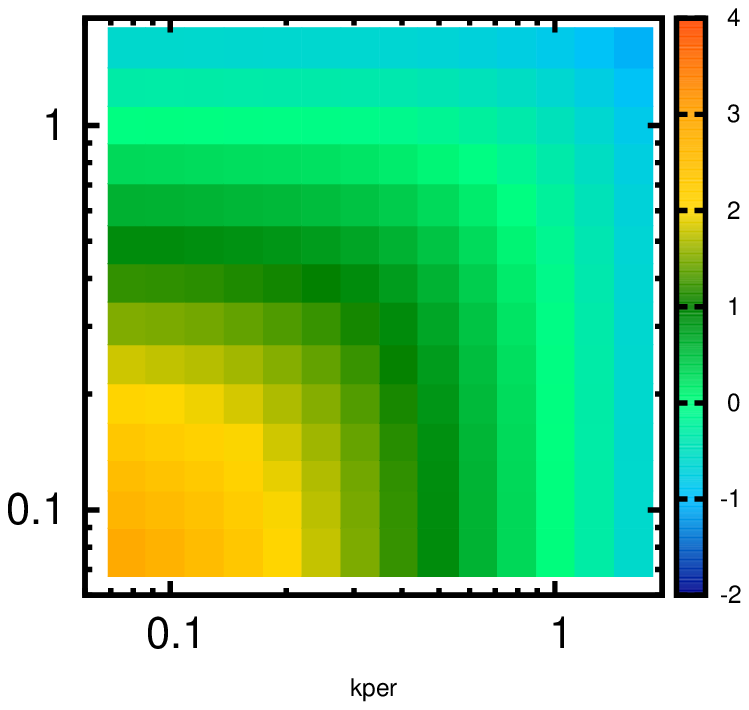}
\hspace*{-2.5cm}
\includegraphics[width=80mm,height=60mm,angle=0]{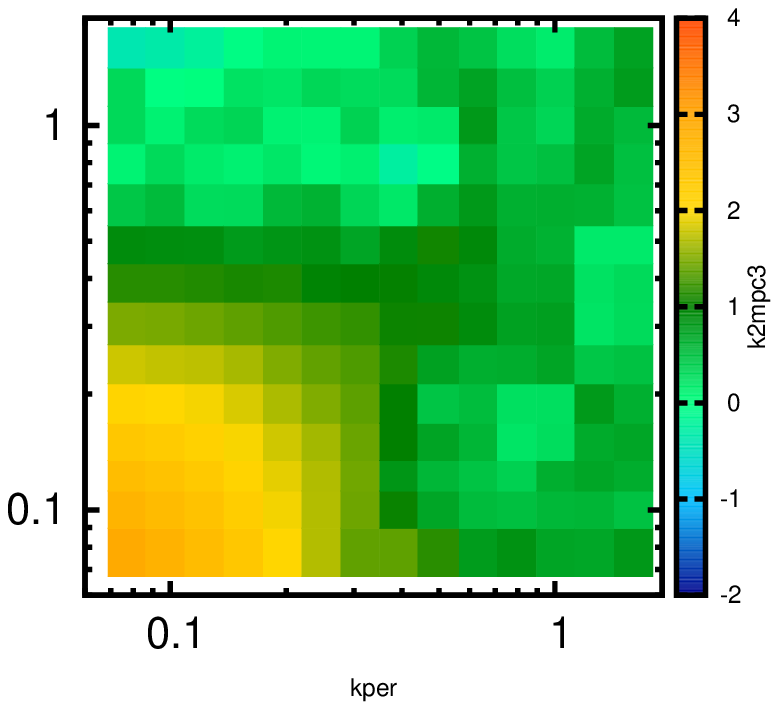}
\caption{This shows the 2D Cylindrical Power Spectrum for
  $n=-3$. The left panel shows the input model power spectrum.  The
  middle and right panels show the estimated power spectrum for
  $f=0.6$ without and with noise respectively.}
\label{fig:fig9}
\end{center}
\end{figure*}

\begin{figure*}
\begin{center}
\psfrag{kpara}[h][h][1.3][0]{$\kp [\rm {Mpc}^{-1}]$}
\psfrag{kper}[b][b][1.3][0]{$\kpm [\rm {Mpc}^{-1}]$}
\psfrag{k2mpc3}[t][t][1][0]{$\rm K^2{Mpc}^3$}
\hspace*{-3.0cm}
\includegraphics[width=80mm,height=60mm,angle=0]{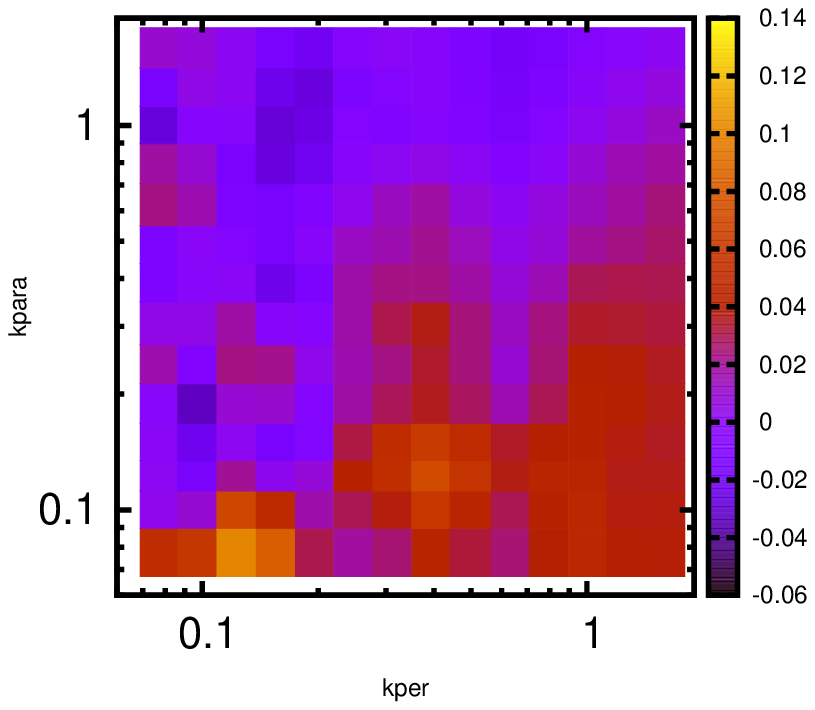}
\hspace*{-2.5cm}
\includegraphics[width=80mm,height=60mm,angle=0]{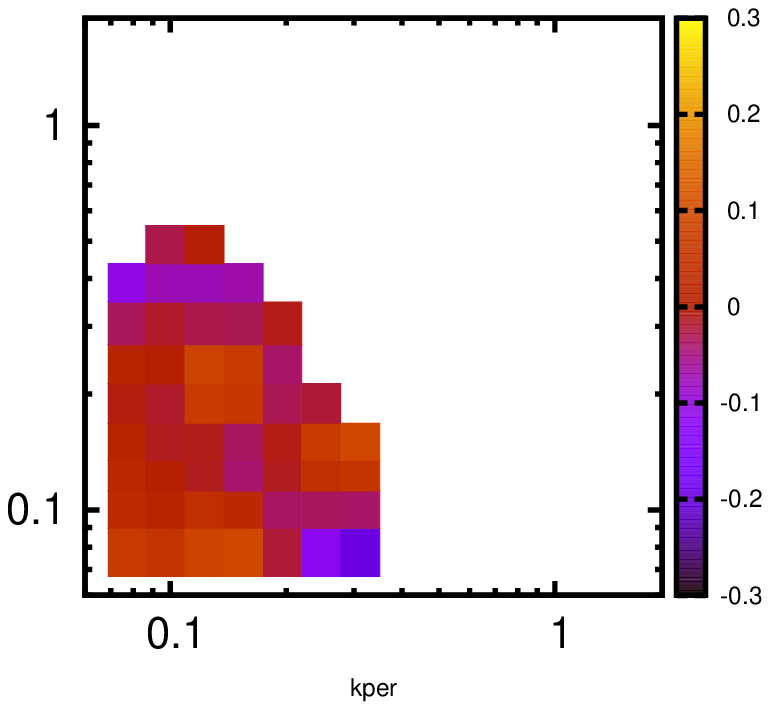}
\caption{The left and right panels show the fractional 
deviation $(P^M(k_\perp,k_\parallel)-P(k_\perp,k_\parallel))/P(k_\perp,k_\parallel)$ 
 without and with noise respectively for $n=-3$ and $f=0.6$.}
\label{fig:fig9a}
\end{center}
\end{figure*}

\begin{figure*}
\begin{center}
\psfrag{kpara}[h][h][1.2][0]{$\kp [\rm {Mpc}^{-1}]$}
\psfrag{kper}[b][b][1.2][0]{$\kpm [\rm {Mpc}^{-1}]$}
\psfrag{k2mpc3}[t][t][1][0]{log($\sige$)}
\includegraphics[width=80mm,angle=0]{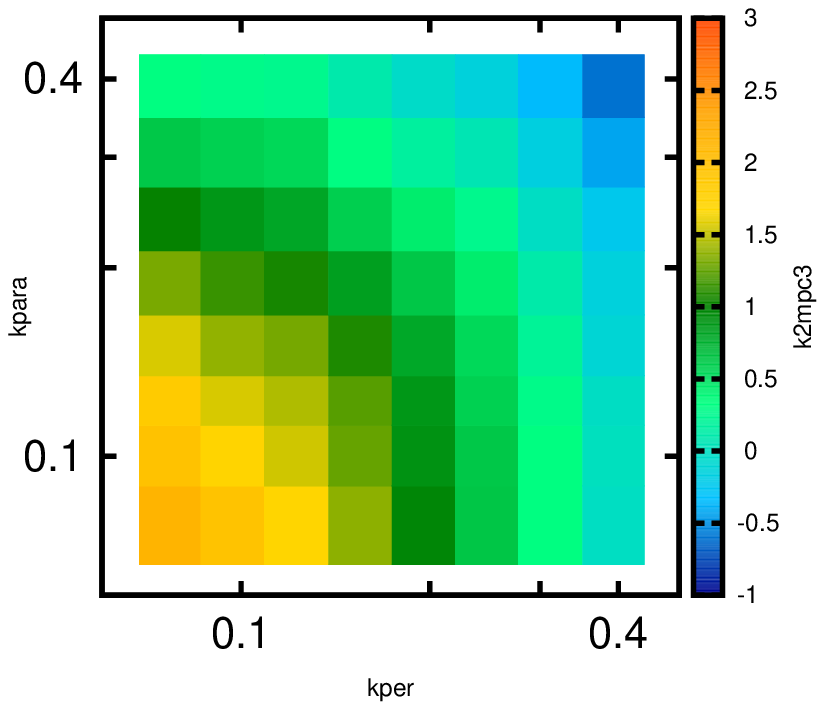}
\includegraphics[width=80mm,angle=0]{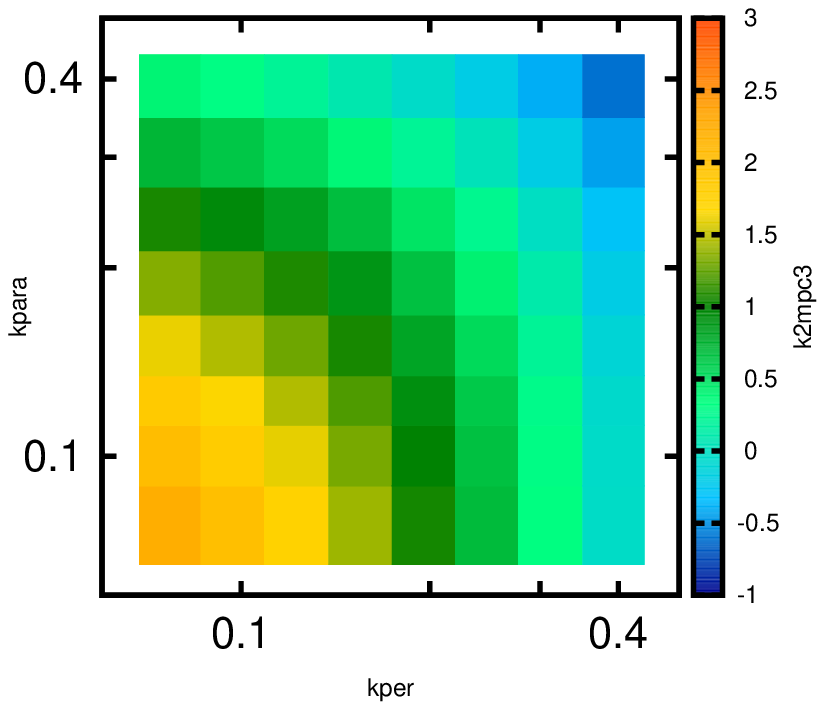}
\includegraphics[width=80mm,angle=0]{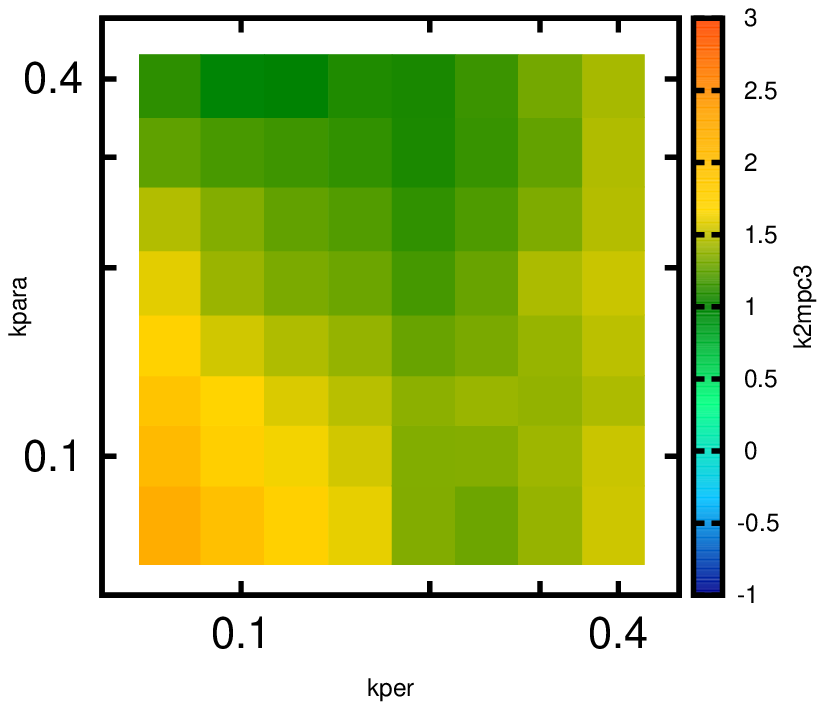}
\includegraphics[width=80mm,angle=0]{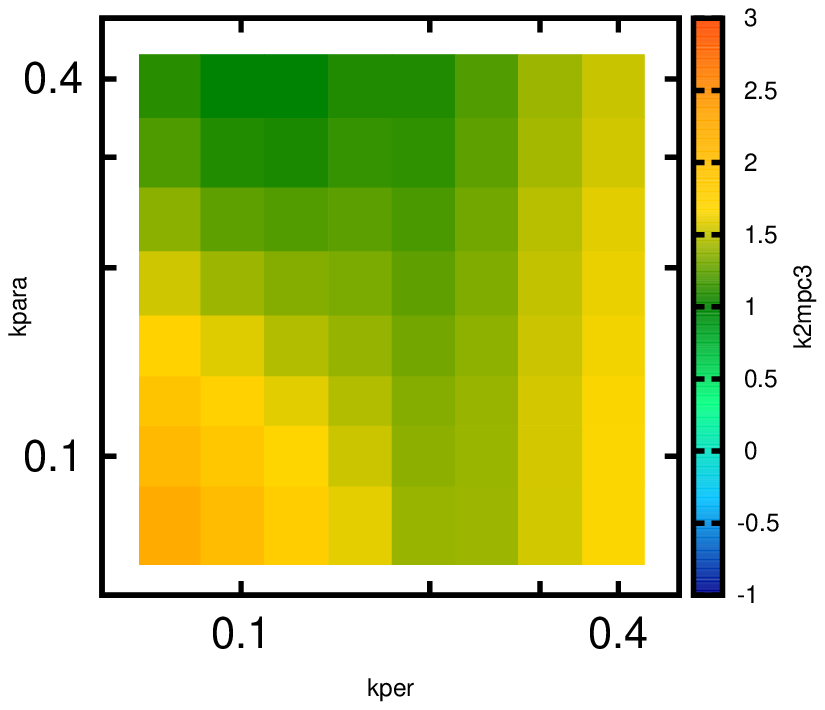}
\caption{This shows the statistical fluctuation ($\sige$) for
 the   2D Cylindrical Power Spectrum for $n=-3$ and $f=0.6$. The upper and
  lower panels show the results without and with system noise
  respectively, the left and right panels show the results from the
  simulations and the analytic prediction respectively.}
\label{fig:fig10}
\end{center}
\end{figure*}

\begin{figure*}
\begin{center}
\psfrag{kpara}[h][h][1.2][0]{$\kp [\rm {Mpc}^{-1}]$}
\psfrag{kper}[b][b][1.2][0]{$\kpm [\rm {Mpc}^{-1}]$}
\psfrag{k2mpc3}[t][t][1][0]{}
\includegraphics[width=80mm,angle=0]{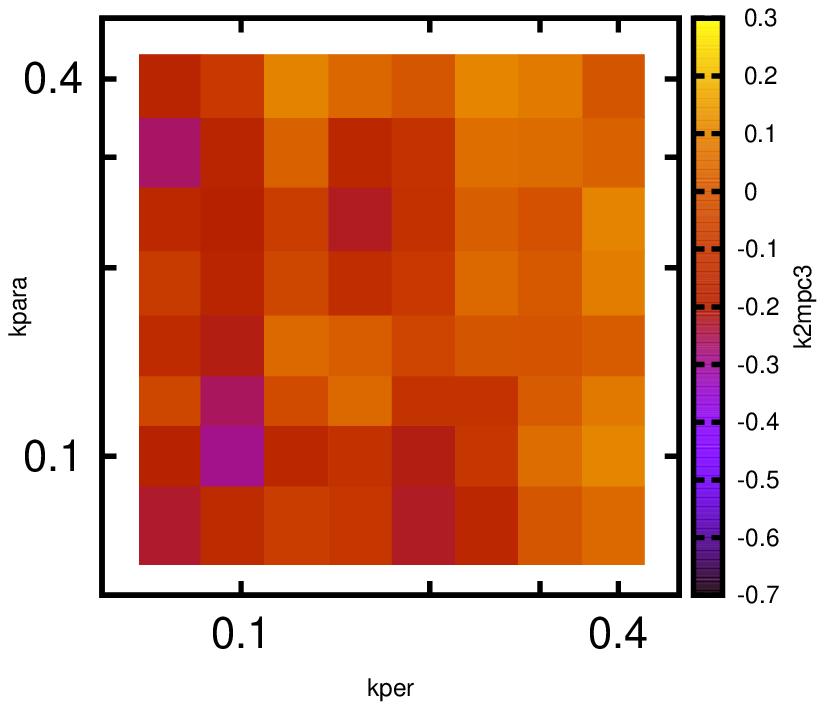}
\includegraphics[width=80mm,angle=0]{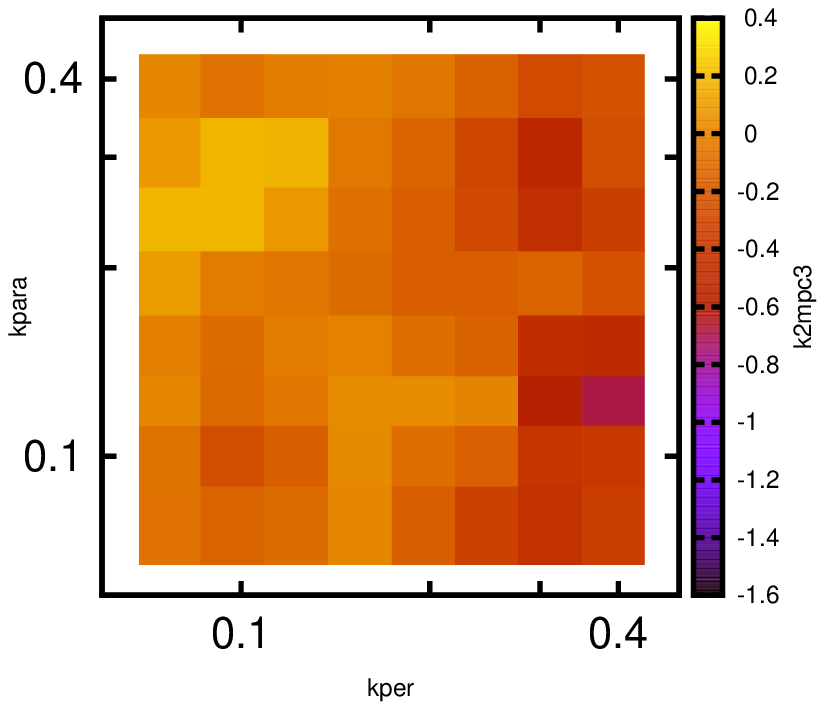}
\caption{The left and right panels  show
    the fractional deviation of $\sige$
    without and with system noise respectively.}
\label{fig:fig10a}
\end{center}
\end{figure*}

We now investigate how well the analytic prediction
(eq. \ref{eq:3dvar4}) for $\sige$ compares with the values obtained
from the simulations (Figure \ref{fig:fig10} ) for $f=0.6$. The two upper
panels consider the situation where there is no system noise for which
the left and right panels respectively show the simulated and the
analytic prediction for the statistical fluctuation $\sige$. We find
that the analytic predictions match quite well with the simulation for
the entire $\mathbf{k}$ range. The two lower panels consider the situation
where the system noise is included for which the left and right panels
respectively show the simulated and the analytic prediction for
$\sige$. The left and right panels of Figure \ref{fig:fig10a} show
  the fractional deviation between the simulated and analytic $\sige$ 
  without and with system noise respectively. We find that 
we have less than $20 \%$ fractional deviation   in $73 \%$ and $64 \%$ of 
the bins in $(k_\perp,k_\parallel)$ space without and with system noise 
respectively. The fractional deviation shows a larger spread in values 
when the system noise is included as compared to the situation without 
system noise. We however do not find any obvious pattern in the distribution
 of the  bins which show a high fractional deviation.

\section{Summary and Conclusions}
Quantifying the statistical properties of the diffuse sky signal directly from
the visibilities measured in  low frequency radio-interferometric  observation
is an important issue.  In this paper we present a statistical estimator,
namely the Tapered Gridded Estimator (TGE), which has been  developed for this 
purpose. The measured visibilities are here  gridded in the $uv$ plane 
to reduce the complexity of the computation. The contribution from the
discrete sources in the periphery of the telescope's FoV,
particularly the sidelobes, pose a problem for power spectrum estimation. The
TGE suppresses the contribution from the outer regions by tapering the
sky response through a suitably chosen window function. The TGE also internally
estimates the noise bias from the input data, and subtracts this out to
give an unbiased estimate of the power spectrum. In addition to the
mathematical formalism for the estimator and its variance, we also present
simulations of  $150 \, {\rm MHz}$ GMRT observations which are used to
validate the estimator.

We have  first considered a situation where we have
observation at a single  frequency for which the 2D TGE provides an 
estimate of  the angular power spectrum $C_{\ell}$. The work here presents an
improvement over an earlier version of the 2D TGE presented in Paper I. 
 This is important in the context of the diffuse Galactic  synchrotron
 emission which is one of the 
major  foregrounds for the cosmological 21-cm signal. Apart from this, the
diffuse  Galactic synchrotron  emission is a probe of the cosmic
ray electrons and the magnetic fields in the ISM of our own Galaxy, and this
is an important study in its own right.

It is necessary to  also include the frequency variation of the sky signal in 
order to quantify the cosmological  21-cm signal. Here the 3D TGE provides an
estimate of $P(\k)$ the power spectrum of the 21-cm brightness temperature
fluctuations. We have considered two different binning schemes which provide
the 1D Spherical Power Spectrum $P(k)$ and the 2D Cylindrical Power Spectrum  
$P(k_\perp,k_\parallel)$  respectively. 
In all cases, we find that the TGE is able to accurately recover the input
model used for  the simulations.  The analytic predictions for the variance are
also found to be in reasonably good agreement with the simulations in most
situations.

Foregrounds are possibly   the biggest challenge for detecting the
cosmological 21-cm power spectrum. Various studies
(eg. \citealt{adatta10}) show that the foreground contribution to the
Cylindrical Power Spectrum $P(k_{\perp},k_{\parallel})$ is expected to
be restricted within a wedge in the $(k_{\perp},k_{\parallel})$ plane.
The extent of this ``foreground wedge'' is determined  by the angular extent 
 of the telescope's FoV. In principle,  it is  possible to 
limit the extent of the foreground wedge by tapering the telescope's FoV. 
In the context of estimating the angular power spectrum $C_{\ell}$, 
our earlier work (Paper II) has demonstrated 
that the 2D TGE is able to suppress the contribution from the outer parts 
and the sidelobes of the telescope's beam pattern. 
We have not explicitly considered the
foregrounds in our analysis of the 3D TGE presented in this paper. We however
expect the 3D TGE to suppress the contribution from the outer parts and the 
sidelobes of the telescopes beam pattern while estimating the power spectrum 
$P(k_{\perp},k_{\parallel})$, thereby reducing the area in the 
$(k_{\perp},k_{\parallel})$ plane  under the foreground wedge.

The 3D TGE holds the promise of allowing us to reduce the extent of the 
foreground wedge by tapering the sky response. It is, however, necessary to 
note that this comes at a cost which we now discuss. First,  we lose 
information  at the largest angular scales due to the reduced FoV. 
This restricts the smallest $k$ value at which it is possible to estimate
the power spectrum. Second,  the reduced FoV results in a larger cosmic 
variance for the smaller angular modes which are within the tapered FoV. 
The actual value of the tapering parameter $f$ that would be
 used to estimate $P(k_{\perp},k_{\parallel})$ will possibly be determined by 
optimising between the cosmic variance and the foreground contribution.  
A possible strategy would be to use different values of $f$  for different 
bins in the  $(k_{\perp},k_{\parallel})$ plane. 
It is also necessary to note that the effectiveness of the tapering 
proposed here depends on the actual baseline distribution, and a 
reasonably dense $uv$ coverage is required for a proper implementation 
of the TGE. We propose to include foregrounds in the simulations 
and address these issues in future work. 
We also plan to apply this estimator to $150 \, {\rm MHz}$ GMRT
data  in future.

\section{Acknowledgements}
S. Choudhuri would like to acknowledge the University Grant Commission, India for providing financial support through Senior Research Fellowship. S. Chatterjee is supported by a University Grants Commission Research Fellowship. SSA would like to acknowledge C.T.S, I.I.T. Kharagpur for the use of its facilities and  thank the authorities of the IUCAA, Pune, India for providing the Visiting Associateship programme.

\end{document}